\PassOptionsToPackage{pdfpagelabels=false}{hyperref} 
\documentclass[fleqn,usenatbib,useAMS]{mnras}
\usepackage{xspace}

\usepackage{graphicx}
\usepackage{inputenc}
\usepackage[Symbol]{upgreek}
\usepackage[T1]{fontenc}
\usepackage{lmodern}
\usepackage{inputenc}
\usepackage{epsfig}
\usepackage{hyperref}
\usepackage{amsmath}
\usepackage{lmodern}
\usepackage[T1]{fontenc}
\usepackage{natbib}
\usepackage{relsize}
\usepackage{ae,aecompl}
\usepackage{times}
\usepackage{subcaption}
\usepackage{graphicx}	
\usepackage{amsmath}	
\usepackage{amssymb}	
\newcommand{\kms}{km~s$^{-1}$ }
\newcommand{\kmsp}{km~s$^{ -1}$}
\newcommand{\oi}{\ion{O}{i}~}

\newcommand{\nv}{\ion{N}{v}~}
\newcommand{\cii}{\ion{C}{ii}~}
\newcommand{\alii}{\ion{Al}{ii}~}
\newcommand{\aliip}{\ion{Al}{ii}}
\newcommand{\aliii}{\ion{Al}{iii}~}
\newcommand{\aliiip}{\ion{Al}{iii}}

\newcommand{\siii}{\ion{Si}{ii}~}
\newcommand{\si}{\ion{Si}{i}~}
\newcommand{\siiii}{\ion{Si}{iii}~}
\newcommand{\sii}{\ion{S}{ii}~}
\newcommand{\siiifs}{Si~{\sc II}$^\ast$~}
\newcommand{\siiv}{\ion{Si}{iv}~}
\newcommand{\vcen}{$v_{\text{cen}}$~}

\newcommand{\oip}{\ion{O}{i}}
\newcommand{\ovip}{\ion{O}{vi}}
\newcommand{\ovi}{\ion{O}{vi}~}

\newcommand{\nvp}{\ion{N}{v}}
\newcommand{\ciip}{\ion{C}{ii}}
\newcommand{\siiip}{\ion{Si}{ii}}
\newcommand{\siiifsp}{Si~{\sc II}$^\ast$}
\newcommand{\siiiip}{\ion{Si}{iii}}
\newcommand{\feii}{\ion{Fe}{ii}~}

\newcommand{\civp}{\ion{C}{iv}}
\newcommand{\civ}{\ion{C}{iv}~}
\newcommand{\siivp}{\ion{Si}{iv}}
\newcommand{\mos}{$\dot{M}_\mathrm{OVI}$}
\newcommand{\mop}{$\dot{M}_\mathrm{ph}$}
\newcommand{\vcenp}{$v_{\text{cen}}$}
\newcommand{\vnp}{$v_{90}$}

\newcommand{\mout}{$\dot{M}$ }

\newcommand{\mstarp}{$M_\ast$}

\newcommand{\moutp}{$\dot{M}$}

\newcommand{\sfr}{M$_\odot$~yr$^{-1}$ }
\newcommand{\sfrp}{M$_\odot$~yr$^{-1}$}

\newcommand{\megasaura}{M\textsc{eg}a\textsc{S}a\textsc{ura}}
\newcommand{\megasauralong}{The Magellan Evolution of Galaxies Spectroscopic and Ultraviolet Reference Atlas}

\begin{document}

\title[Feeding the fire of 10$^7$~K galactic outflows]{Feeding the fire: Tracing the mass-loading of 10$^7$~K galactic outflows with \ovi absorption}
\author[Chisholm et al.]{J. Chisholm$^{1}$\thanks{Contact email: John.Chisholm@unige.ch}, R. Bordoloi$^{2}$\thanks{Hubble Fellow}, J.R. Rigby$^{3}$ and M. Bayliss$^{2}$\\
$^{1}$  Observatoire de Gen\`{e}ve, Universit\'{e} de Gen\`{e}ve, 51 Ch. des Maillettes, 1290 Versoix, Switzerland\\
$^{2}$ MIT Kavli Institute for Astrophysics and Space Research, 77 Massachusetts Ave., Cambridge, MA 02139, USA\\
$^{3}$ Observational Cosmology Lab, NASA Goddard Space Flight Center, 8800 Greenbelt Rd., Greenbelt, MD 20771, USA\\
}
\pubyear{2017}
\label{firstpage}
\pagerange{\pageref{firstpage}--\pageref{lastpage}}
\maketitle
\begin{abstract}
Galactic outflows regulate the amount of gas galaxies convert into stars. However, it is difficult to measure the mass outflows remove because they span a large range of temperatures and phases. Here, we study the rest-frame ultraviolet spectrum of a lensed galaxy at $z \sim 2.9$ with prominent interstellar absorption lines from \oip, tracing neutral gas, up to \ovip, tracing transitional phase gas. The \ovi profile mimics weak low-ionization profiles at low velocities, and strong saturated profiles at high velocities. These trends indicate that \ovi gas is co-spatial with the low-ionization gas. Further, at velocities blueward of -200~\kms the column density of the low-ionization outflow rapidly drops while the \ovi column density rises, suggesting that \ovi is created as the low-ionization gas is destroyed. Photoionization models do not reproduce the observed \ovip, but adequately match the low-ionization gas, indicating that the phases have different formation mechanisms. Photoionized outflows are more massive than \ovi outflows for most of the observed velocities, although the \ovi mass outflow rate exceeds the photoionized outflow at velocities above the galaxy's escape velocity. Therefore, most gas capable of escaping the galaxy is in a hot outflow phase. We suggest that the \ovi absorption is a temporary by-product of conduction transferring mass from the photoionized phase to an unobserved hot wind, and discuss how this mass-loading impacts the observed circum-galactic medium.
\end{abstract}

\section{INTRODUCTION}

Supernovae, stellar winds, cosmic rays and high-energy photons from high-mass stars accelerate gas out of star-forming regions as a galactic outflow \citep{heckman90, heckman2000, veilleux05}. Removing gas from galaxies stops runaway star formation by controlling the amount of gas converted into stars \citep{larson74, white91, hopkins14}, which may establish the observed star formation history of the universe \citep{madau, springel03, oppenheimer06}.
\\\

While studies have established that outflows are ubiquitous in star-forming galaxies at all redshifts \citep{heckman90, heckman2000, pettini2002, veilleux05, martin2005,rupke2005b, weiner, rubin13, bordoloi14, heckman2015, bordoloi16a, chisholm16}, the amount of mass ejected by outflows--the mass outflow rate--is challenging to constrain. Galactic outflows are diffuse, making them difficult to observe with emission lines outside of the~local~universe \citep{westmoquette, sharp, arribas2014}. Typically, studies use metal absorption lines to probe the gas along the line-of-sight, which provides kinematic and column density information. However, converting these measurements into a mass outflow rate requires assumptions about the geometry, ionization structure, and metallicity of the outflow. These assumptions add up to a factor of ten uncertainty to most observed mass outflow rates \citep{murray07, chisholm16}.

Most galactic outflow studies focus on warm gas at approximately 10$^4$~K because this gas phase has strong absorption and emission lines in the rest-frame ultraviolet and optical. However, galactic outflows are multiphase phenomena. Dense molecular outflows are observed in the local universe \citep{weiss99, matsushita2000, leroy15, walter2017}, with molecular mass outflow rates comparable to the neutral mass outflow rate, and up to 1.3 times larger than the star formation rate \citep{chisholmmat}. At the other end of the temperature range, X-ray observations of low redshift galaxies reveal that a pervasive large-scale wind fluid, with a temperature greater than 10$^{7}$~K, dominates the thermal and kinetic energy of outflows \citep{griffiths, strickland2000, strickland09}. These molecular and hot winds may contain at least as much mass and energy as the 10$^4$~K phase; studies focusing only on warm outflows likely underestimate the total mass outflow rate. 

Determining the mass outflow rate is additionally complicated by interactions between the different outflowing phases. Phase transitions convert molecular gas to atomic \citep{leroy15}, and photoionized gas is further ionized into the hot wind fluid \citep{strickland2000, chisholm16b}, producing a shifting ionization structure. Consequently, a given outflow phase can dominate the mass and energy budget of an outflow at a given radius, velocity, or time. \citet{strickland2000} find that most of the soft X-rays from the local starburst M~82 are produced when mass transfers from cooler phases into the hotter wind. These regions are called mass-loading regions because they feed most of the mass into the hot wind fluid \citep{maclow, suchkov, strickland09}.

The hot wind can only be probed by X-ray observations, and only the nearest starbursts are bright enough to be observed in X-rays. Alternatively,  the hot wind can be studied by observing gas that is temporarily created during mass-loading. With a cooling time of 1~Myr, "transitional" gas is short-lived because it sits near the peak of the cooling curve \citep[$\sim5 \times 10^5$~K;][]{spitzer56} and it must be observed shortly after it is created. The \ovip~1032\AA\ doublet  is one of the few observable tracers of this transitional gas.

Observations of the \ovi doublet are challenging because the transition is in the rest-frame far-ultraviolet (FUV), and must be observed from space for local galaxies. Complicating the situation, the high resolution configuration of the Cosmic Origins Spectrograph on the {\sl Hubble Space Telescope} cannot observe the \ovi doublet in local galaxies. The Far Ultraviolet Spectroscopic Explorer (FUSE) studied the \ovi region in 16 galaxies, finding the \ovi absorption line to have a velocity centroid that is 30~\kms more blueshifted than warm gas tracers, like \ciip, but with a similar maximum velocity \citep{heckman01, grimes06, grimes07, grimes2009}. Few other studies have observed this important outflow phase, and the origin of \ovi outflows is still unsettled.   

In this paper, we analyze new spectroscopic observations of the \ovip~1032\AA\ absorption line from a gravitationally lensed, high redshift galaxy, SGAS~J122651.3+215220, that is drawn from the Magellan Evolution of Galaxies Spectroscopic and Ultraviolet Reference Atlas (\megasaura). The moderate resolution \megasaura\ spectra provide unprecedented far-ultraviolet wavelength coverage, with absorption lines spanning a factor of 17 in ionization potential, allowing for a detailed comparison of the \ovi line profile to low-ionization lines. First, we summarize the data reduction (\autoref{data}) and analysis of the absorption lines (\autoref{lines}). Then, we compare the velocity-resolved \ovi line profile to the low-ionization absorption lines (\autoref{profile}). In \autoref{cloudy models}, we model the low-ionization lines with photoionization models to determine whether photoionization models reproduce the observed \ovip. Finally, we discuss: \ovi formation mechanisms (\autoref{o6}), upper limits for the column density and the mass outflow rate of each phase (\autoref{phases}), and the physical picture suggested by these observations (\autoref{model}). With this analysis we explore what feeds the elusive and powerful 10$^7$~K hot wind. 
\\\

\section{\megasaura\ DATA}
\label{data}

\begin{figure}
\includegraphics[width = 0.5\textwidth]{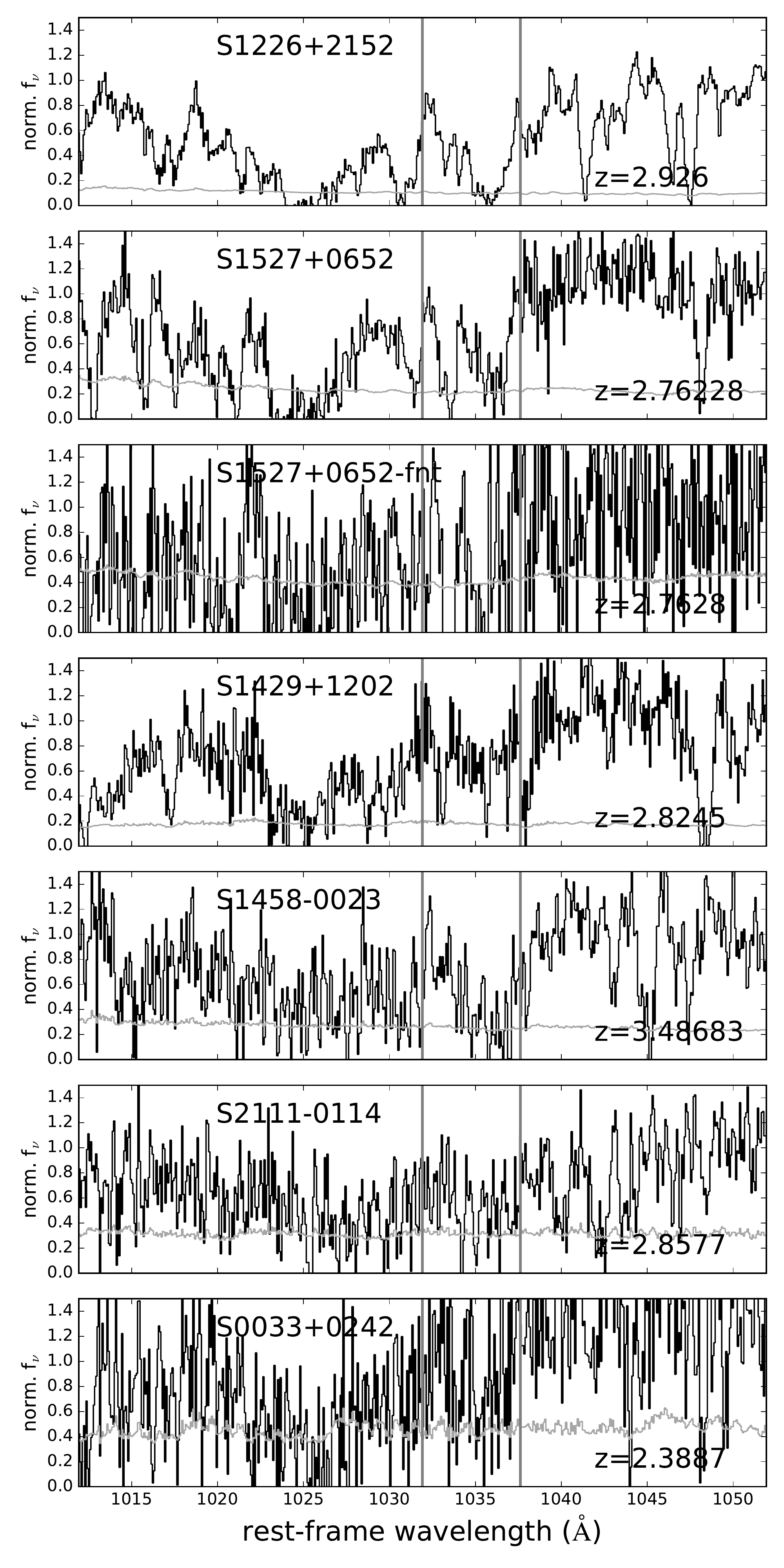}
\caption{The \ovi 1032\AA\ region for six galaxies from the \megasaura\ sample, arranged in descending order of their signal-to-noise ratio at the \ovi 1032\AA\ line, expect for J1527+0652-fnt which is a fainter knot from J1527+0652. The \ovi 1032\AA\ and 1038\AA\ line are both marked by grey vertical lines. \ovip~1032\AA\ absorption is found in three of the four highest signal-to-noise observations (J1226+2152, J1527+0652, and J1458-0023).}
\label{fig:megasaura}
\end{figure}

The spectra used in this paper are drawn from Project \megasaura: \megasauralong\ \citep{rigbya}. In short, some of the brightest gravitational lenses in the sky were observed with the Magellan Echellette (MagE) Spectrograph \citep{Marshall:2008bs} on the Magellan telescopes. The highest redshift galaxies within the \megasaura\ sample have MagE observations with rest-frame ultraviolet wavelengths between $850 \la \lambda_r \la 2200$~\AA, suitable to analyze the \ovi absorption line. The spectra were reduced using D.~Kelson's Carnegie Python pipeline\footnote{The pipeline can be found at \url{http://code.obs.carnegiescience.edu/mage-pipeline}}, and were corrected for Milky Way foreground reddening using the observed extinction from Pan-STARRS and 2MASS observations \citep{green15}, assuming the Milky Way extinction law \citep{cardelli}.
\\

In \autoref{fig:megasaura} we show the rest-frame spectra of the six \megasaura\ galaxies with  \ovip~1032\AA\ coverage. With the exception of J1429+1202, galaxies with SNR$~>~3$ show broad \ovip~1032\AA\ absorption that is blueshifted from the stellar continuum, as demonstrated by the grey lines in \autoref{fig:megasaura}. In a future paper, we will explore the outflow properties of the entire \megasaura\ spectral data set, but here we focus on a single galaxy, SGAS~J122651.3+215220 \citep[hereafter J1226+2152;][]{koester}, because the high signal-to-noise ratio ($\sim10$ at 1032\AA) allows us to clearly define and characterize the \ovi line profile. J1226+2152 has a redshift of $2.926\pm0.0002$, as measured from the [\ion{C}{iii}]~1907\AA\ and \ion{C}{iii}]~1909\AA\ nebular emission doublet \citep{rigbya}. The total integration time for J1226+2152 was 12.42~hr, spread over multiple observing runs, with a final spectral resolution, as measured from night sky emission lines, of $R=4010 \pm 170$ ($75\pm3$~\kmsp). The median signal-to-noise ratio per resolution element was $SNR=20$.

\section{SPECTRAL ANALYSIS}
\label{lines}
\subsection{Continuum fitting}
\label{cont}
\begin{figure*}
\includegraphics[width = \textwidth]{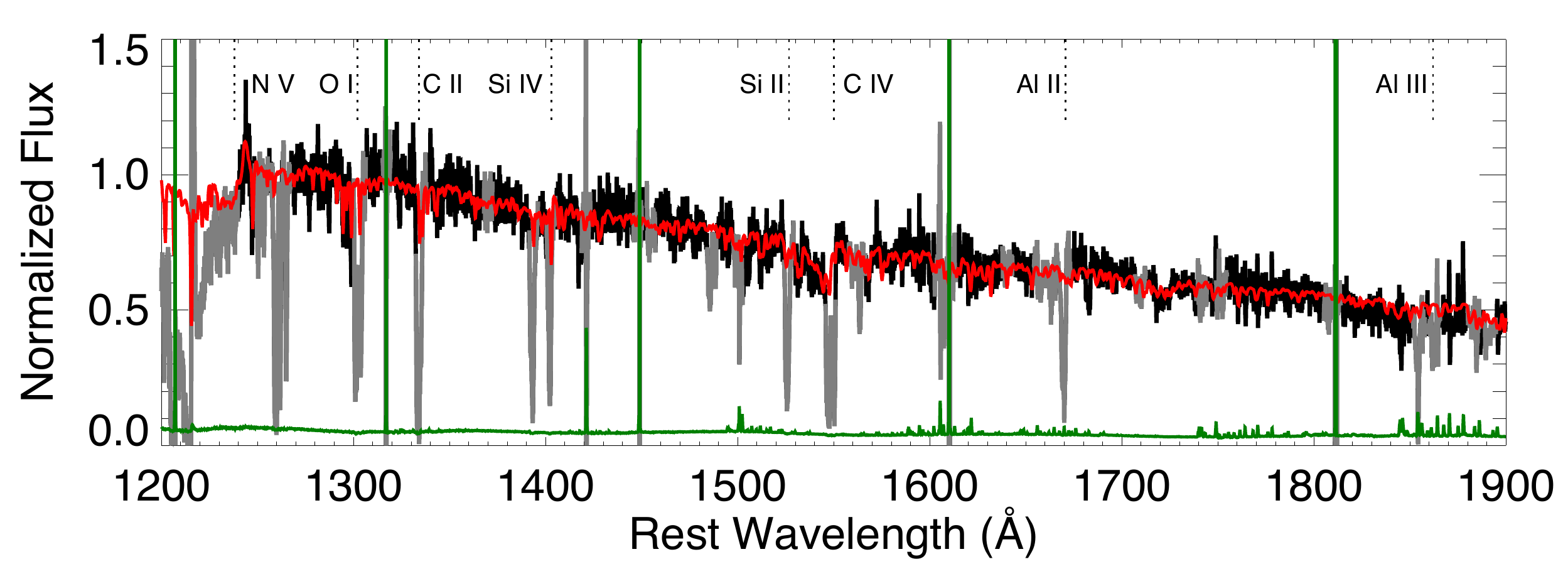}
\caption{Observed spectrum (in F$_\lambda$) of J1226+2152 in black for the restframe wavelength regime between 1200--1900\AA, with the error array included in dark green. Gray areas are masked out of the stellar continuum fitting to avoid interstellar and intervening absorption lines. Over-plotted in red is the {\small STARBURST99} stellar continuum fit to the data (see \autoref{cont}). The stellar continuum fit establishes the continuum level and the zero-velocity (systematic) of the spectrum. The best fit stellar continuum model has a light-weighted age of 11~Myr, a stellar metallicity of 0.2~Z$_\odot$, and stellar extinction of E(B-V) = 0.15~mag. Interstellar absorption lines used in the paper are marked by dashed lines in the upper portion of the plot.}
\label{fig:cont_red}
\end{figure*}
\begin{figure}
\includegraphics[width = 0.5\textwidth]{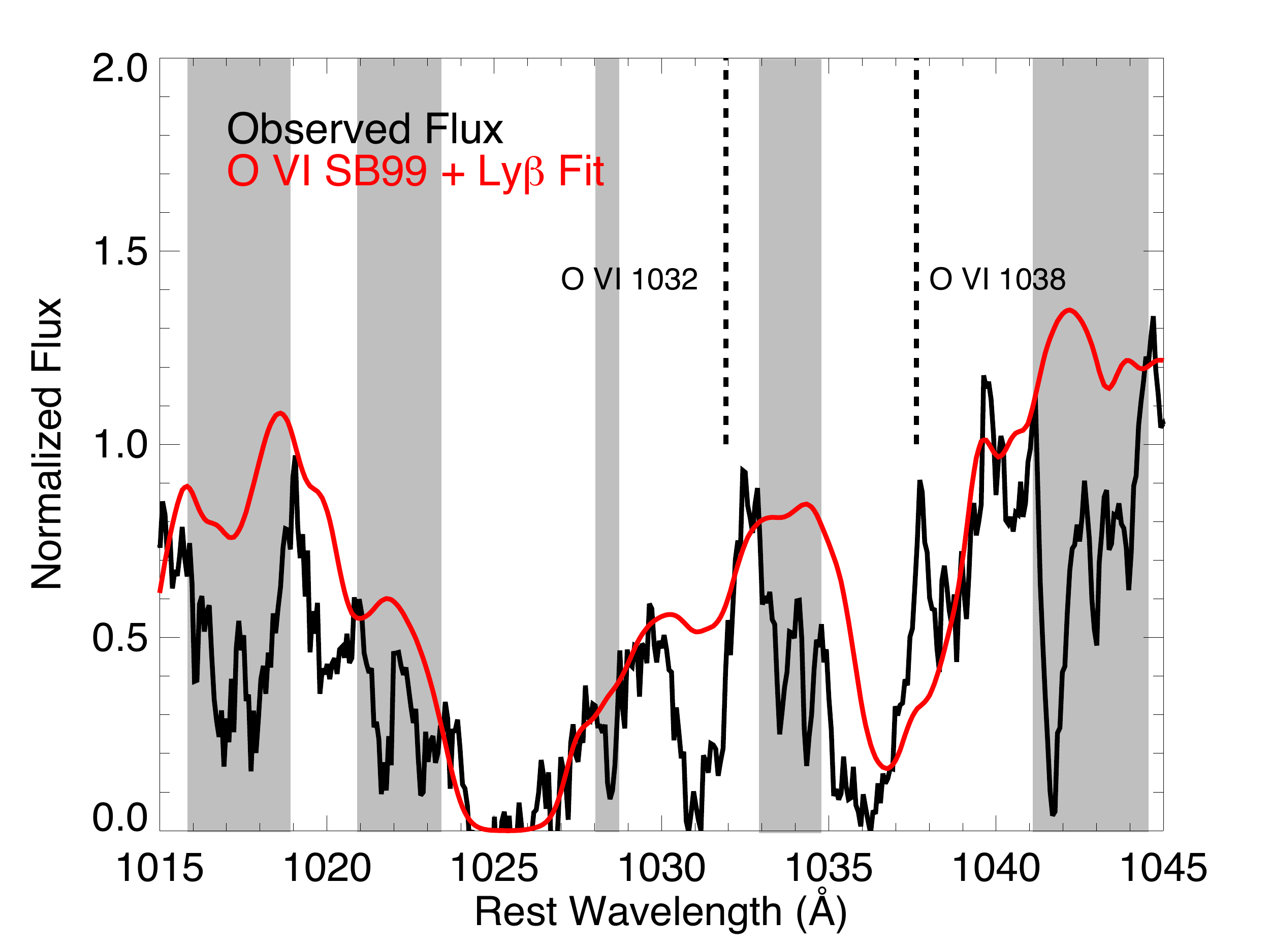}
\caption{The simultaneous {\small STARBURST99} and Ly$\beta$ fit (red line) to the \ovi region (black line) for the lensed galaxy J1226+2152. The zero-velocity, as determined by the redshift from the stellar continuum fitting, of the \ovi 1032\AA\ and 1038\AA\ lines are denoted by dashed vertical lines. The \ovi 1038\AA\ line is heavily blended by nearby interstellar \oi and \cii lines, and we do not use that line in our analysis. The {\small STARBURST99} fit accounts for  the stellar continuum and Ly$\beta$ contributions to establish the unity flux level of the \ovi profile below. Gray regions show areas of intervening absorbers that were masked from the continuum fit.}
\label{fig:cont}
\end{figure}
\begin{figure*}
\includegraphics[width = 0.95\textwidth]{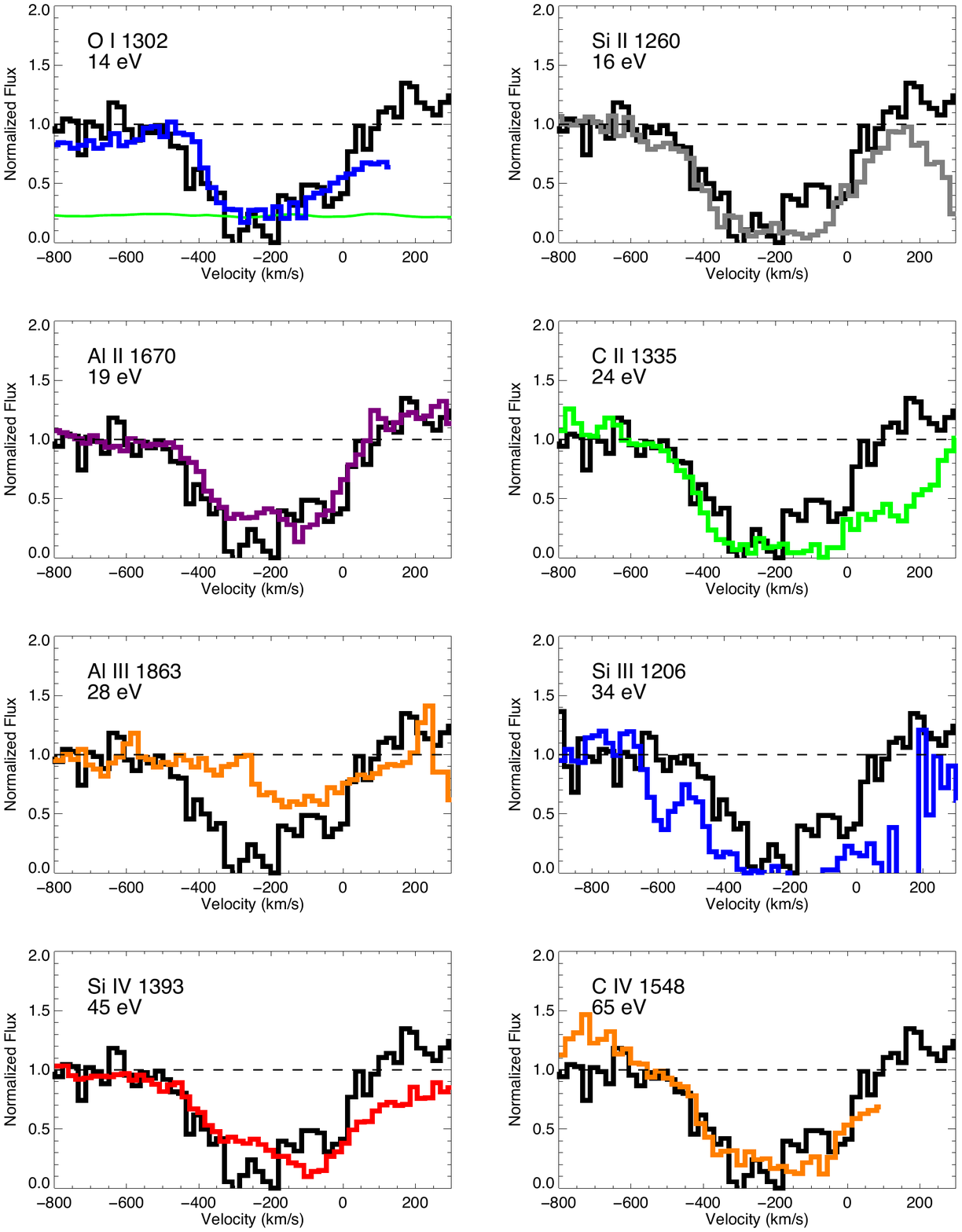}
\caption{Comparison of the stellar continuum-normalized absorption profiles for various ISM metal absorption lines (in colors) with the \ovi line profile in black in all the panels. The panels are ordered in terms of increasing ionization potential of the low-ionization transition. \oip~1302\AA, \aliip~1670\AA, \aliiip~1863\AA, and \siivp~1393\AA\ are not black at any portion of their line profiles, and are considered weak lines. Conversely, the \siiip~1260\AA, \siiiip~1206\AA\ and \ciip~1335\AA\ lines are nearly black over 400~\kmsp, and are considered strong lines. Note that the \siiiip~1206~\AA\ line is close to the broad Ly$\alpha$ absorption line, and we fit a spline to further normalize the spectrum. The error array corresponding to the \ovi region is shown in green in the top left panel (the \oi panel). Blueward of $-200$~\kms the \ovi profile is similar to the strong low-ionization absorption lines--like \siiip, \siiii and \ciip--while the depth and shape of the \ovi profile redward of $-200$~\kms is similar to the weak lines--like \aliip, \aliii and \oip.}
\label{fig:lineprof}
\end{figure*}

The spectrum of J1226+2152 is observed \lq{}\lq{}down-the-barrel\rq{}\rq{} such that  the ISM absorption lines are imprinted onto the stellar continuum. In this subsection we discuss how we fit and remove the stellar continuum to account for the shape and absorption lines of the background stars. We fit the stellar continuum twice: once for wavelengths greater than 1240\AA, and once for wavelengths around the \ovi doublet (1015-1045\AA). The continuum must be fit twice because there are numerous intervening absorbers in the Ly$\alpha$ forest which can drive the stellar continuum fits if they are not properly masked.

Following \citet{chisholm15}, we fit the stellar continuum redward of 1240\AA\ using {\small MPFIT} \citep{mpfit} with a linear combination of 10 single age (with ages between 1-40~Myr), fully theoretical {\sc STARBURST99} models \citep{claus99, claus2010} that use the Geneva stellar evolution models with high mass loss rates \citep{geneva94}, while simultaneously fitting for the stellar continuum extinction (E$_\mathrm{s}$(B-V)) using a Calzetti extinction law \citep{calzetti}. We allow for the five {\sc STARBURST99} stellar continuum metallicities (0.05, 0.2, 0.4, 1.0, 2.0~Z$_\odot$), and choose the best-fit model using a $\chi^2$ test. The best-fit model is over-plotted on the observed data in \autoref{fig:cont_red} and has an E$_\mathrm{S}$(B-V) of $0.15\pm0.002$~mag, a stellar metallicity of 0.2~Z$_\odot$, and a light-weighted age of 11~Myr. We cross-correlate the best-fit model with the data to determine the velocity offset between the observed stellar continuum and the {\sc STARBURST99} models. This velocity offset defines the zero velocity of the spectrum, also known as the systemic redshift. Further, we use the fitted {\sc STARBURST99} model as the ionizing source for the {\small CLOUDY} models in \autoref{cloudy}.

The \ovi doublet, at wavelengths of 1032\AA\ and 1038\AA, is in an extremely complicated spectral region. \ovi in the atmosphere of hot massive stars produces strong P-Cygni profiles \citep{claus99}, while there is strong Ly$\beta$~1026\AA\ absorption from both the background stars and the foreground interstellar medium. Additionally, there are many strong interstellar metal absorption lines, such as \siii 1020\AA, \oi 1025\AA, \cii 1036\AA, and \oi 1039\AA, that blend with the \ovi profiles. Determining the continuum in this region is challenging, but without a physically motivated continuum the broad Ly$\beta$ or stellar \ovi P-Cygni profiles may be mistaken for high-velocity \ovi (see \autoref{fig:cont}). 

We fit the \ovi continuum region of J1226+2152 using a simultaneous model of empirical {\sc STARBURST99} models and a Ly$\beta$ Voigt profile. We mask, by hand, regions of possible intervening absorbers (gray regions in \autoref{fig:cont}) and interstellar metal lines. For simplicity, and because the Calzetti law is not defined at these wavelengths, we do not account for reddening. The resultant fit is shown by the red line in \autoref{fig:cont}, where the zero-velocity of the \ovi doublet is marked by two vertical dashed lines. The weaker \ovi 1038\AA\ line is severely blended by the interstellar \ciip~1036\AA\ and \oip~1039\AA\ absorption lines; we cannot use the \ovip~1038\AA\ line for this reason.

\subsection{Comparing the \ovi absorption profile to low-ionization profiles}
\label{lineprof}
\begin{table*}
\caption{Quantities derived from the interstellar metal absorption lines of J1226+2152. Column 1 indicates the transition, column 2 gives the ionization potential, column 3 gives the velocity at half the equivalent width (\vcenp), column 4 gives the velocity at 90\% of the continuum on the blue portion of the profile (\vnp), column 5 gives the rest-frame equivalent width (W$_r$), column 6 gives the measured apparent optical depth column density calculated using \autoref{eq:colden} ($N$), and column 7 gives the column density of the best-fit {\small CLOUDY} model. Transitions that are blended with other transitions are denoted by a \lq{}\lq{}b\rq{}\rq{}. Possibly saturated lines are marked with an \lq{}\lq{}s\rq{}\rq{}. The \nv 1243\AA\ line is reported as an upper limit because the line is not detected (ND; see \autoref{fig:nvlineprof}). Transitions used in the {\small CLOUDY} modeling are denoted with a \lq{}\lq{}c\rq{}\rq{}. }
\begin{tabular}{lcccccc}
\hline
Line & Ionization Potential & \vcenp & \vnp & W$_r$ & log($N$) & {\sc CLOUDY} log($N$)  \\
 & [eV] & [\kmsp] & [\kmsp] & [\AA]  & [log(cm$^{-2}$)] & [log(cm$^{-2}$)]   \\
  (1) & (2) & (3) & (4) & (5) & (6) & (7) \\
 \hline
\si 1845.51 & 8.2 & $-16 \pm 117$ & $-35 \pm 10$ &  $0.24 \pm 0.03$ & $13.28 \pm 0.13$ & 10.82 \\
\oi 1302.17 & 13.6 & $-175 \pm 8$ & $-429 \pm 14$ & $1.32 \pm 0.05$ & $15.62 \pm 0.11$ & 16.13$^\mathrm{c}$ \\
\feii 1608.46 & 16.2 & $-158 \pm 7$ & $-363 \pm 36$ & $0.64 \pm 0.06$  & $14.85 \pm 0.03$ & 14.34 \\ 
  \siii 1808.00 & 16.3 & $-93 \pm 31$ & $-168 \pm 4$& $0.25 \pm 0.08$ & $15.57 \pm 0.14$ & $15.13$ \\
\siii 1304.37 & 16.3 & $-175 \pm 5^b$ & $^{b}$ & $1.18 \pm 0.12^{b}$ & $15.23 \pm 0.02^{b}$ & 15.13 \\
\siii 1526.71 & 16.3 & $-162 \pm 12$ & $-466 \pm 9$& $1.63 \pm 0.16$ & $15.15 \pm 0.08$ & $15.13^\mathrm{c}$\\
\siii 1190.42 & 16.3 & $-236 \pm 4^b$ & $-631\pm17^{b}$ & $1.91 \pm 0.19^b$ & $>15.0^{s}$ & 15.13\\
\siii 1193.29 & 16.3 & $-169 \pm 4$ & $-450\pm17$ & $1.50 \pm 0.15$ & $>15.0^{s}$ & 15.12 \\
\siii 1260.42 & 16.3 & $-201 \pm 10$ & $-585 \pm 40$ & $1.78 \pm 0.25$ & $>14.46^{s}$ & 15.13 \\
\alii 1670.79 & 18.8& $-196 \pm 14$ & $-448 \pm 8$ & $1.51 \pm 0.15$ & $13.88 \pm 0.04$ & 13.81$^\mathrm{c}$\\
\sii 1250.58 & 23.3 & $-110 \pm 49$ & $-127 \pm 7$ & $0.24 \pm 0.03$ & $15.61 \pm 0.05$ & 14.95 \\
\cii 1334.53 & 24.4 & $-125 \pm 8$ & $-505 \pm 33$ & $2.43 \pm 0.24$ & $>15.52^{s}$ & 16.37\\
\aliii 1854.72 & 28.4 & $-109 \pm 25$ & $-369 \pm 9$ & $1.12 \pm 0.11$ & $14.06 \pm 0.06$ &  14.91\\
\aliii 1862.79 & 28.4 & $-104 \pm 31$ & $-254 \pm 13$ & $0.80 \pm 0.08$ & $14.15 \pm 0.04$ &  $13.91^\mathrm{c}$\\
\siiii 1206.51 & 33.5 & $-184 \pm 98$ & $-650 \pm 13$ & $2.80 \pm 0.28$ & $>14.40^{s}$ & 15.39\\
\siiv 1393.76 & 45.1 & $-130 \pm 9$ & $-455 \pm 10$ & $1.85 \pm 0.18 $ & $14.68 \pm 0.03$ &  14.78\\
\siiv 1403.77 & 45.1 & $-153 \pm 11$ & $-394 \pm 15$ & $1.15 \pm 0.12$ & $14.69 \pm 0.04$ & 14.78$^\mathrm{c}$\\
\civ 1548.20 & 64.5 & $-194 \pm 55^{b}$ & $-478 \pm 150$ & $^{s,b}$ & $>15.10^{s,b}$ & 15.60\\ 
\civ 1550.77 & 64.5 & $-146 \pm 18^{b}$ &  $^{b}$ & $^{s,b}$ & $> 15.23^{s,b}$ & 15.60\\
\nv 1242.80 & 97.9 & $--$ &$--$ & ND &  $<14.25$ & 10.89 \\
\ovi 1031.91
& 138 & $-246 \pm 11$ & $-532 \pm 2$ & $2.46 \pm 0.25$ & $15.35 \pm 0.05^s$ & 0   \\
\end{tabular}
\label{tab:lines}
\end{table*}
\begin{figure}
\includegraphics[width =0.5 \textwidth]{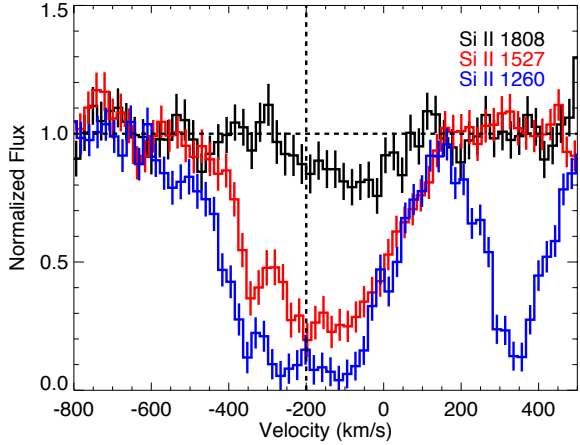}
\caption{The \siiip~1808\AA\ (black), \siiip~1527\AA\ (red), and \siiip~1260\AA\ (blue) absorption profiles. The error bars show the 1$\sigma$ flux errors. The \siii absorption lines have $f$-values of 0.0025, 0.13, and 1.22, respectively. There is an intervening absorber (detected with a doublet finding algorithm)  at $+350$~\kms from the \siiip~1260\AA\ line.  The marginal detection (3$\sigma$) of \siiip~1808\AA\ leads to a \siii column density of log(N[cm$^{-2}$])$ = 15.57\pm0.14$, consistent, within $2\sigma$, with the \siii column densities measured from the \siiip~1527\AA\ line. The weak \siiip~1808\AA\ line is not significantly detected blueward of $-200$~\kms (vertical dashed line), which is the velocity where the photoionized density begins to rapidly decline (see \autoref{proffits}).}
\label{fig:si2}
\end{figure}
\begin{figure}
\begin{subfigure}{\textwidth}
\includegraphics[width = 0.5\textwidth]{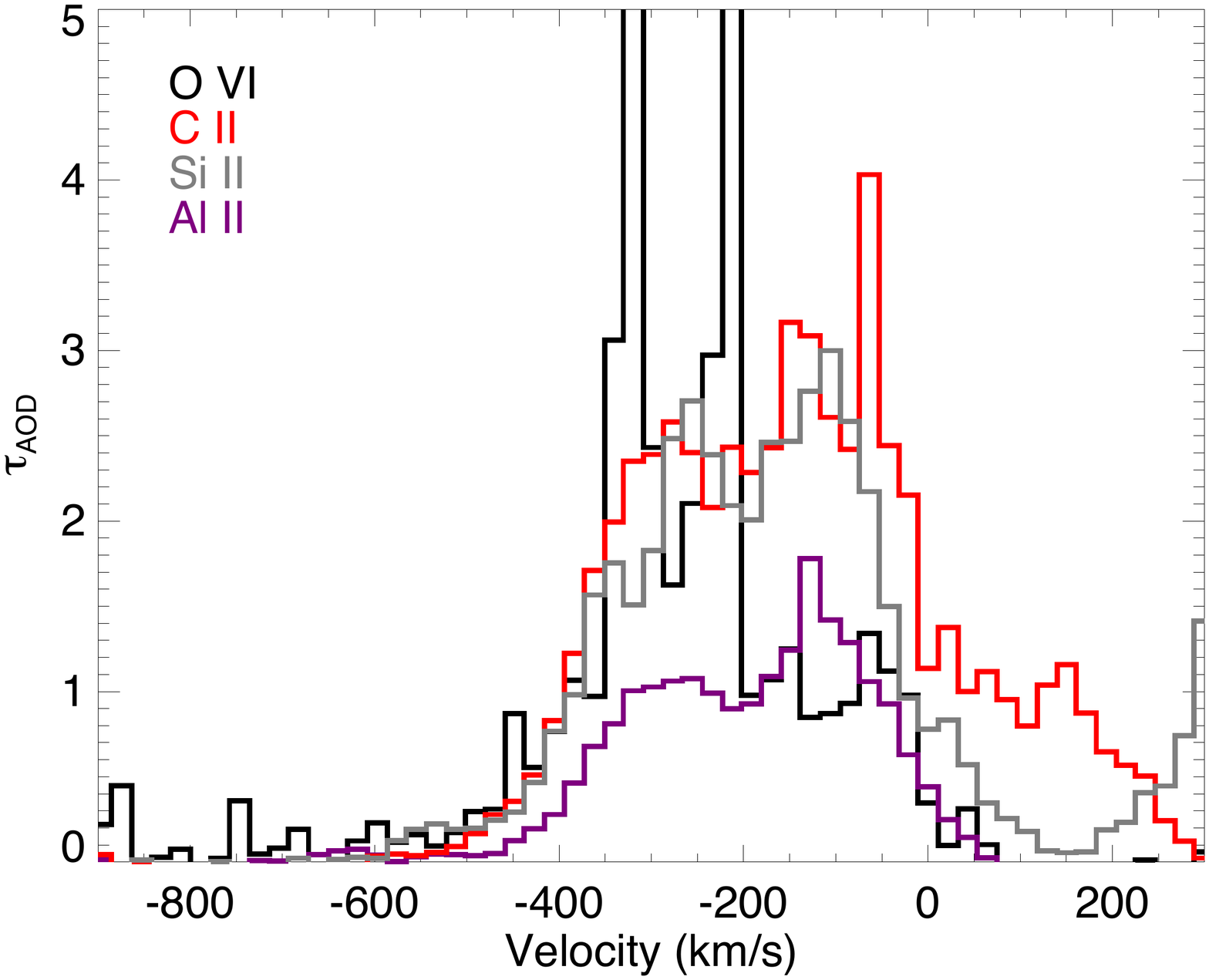}
\end{subfigure}
\begin{subfigure}{\textwidth}
\includegraphics[width = 0.5\textwidth]{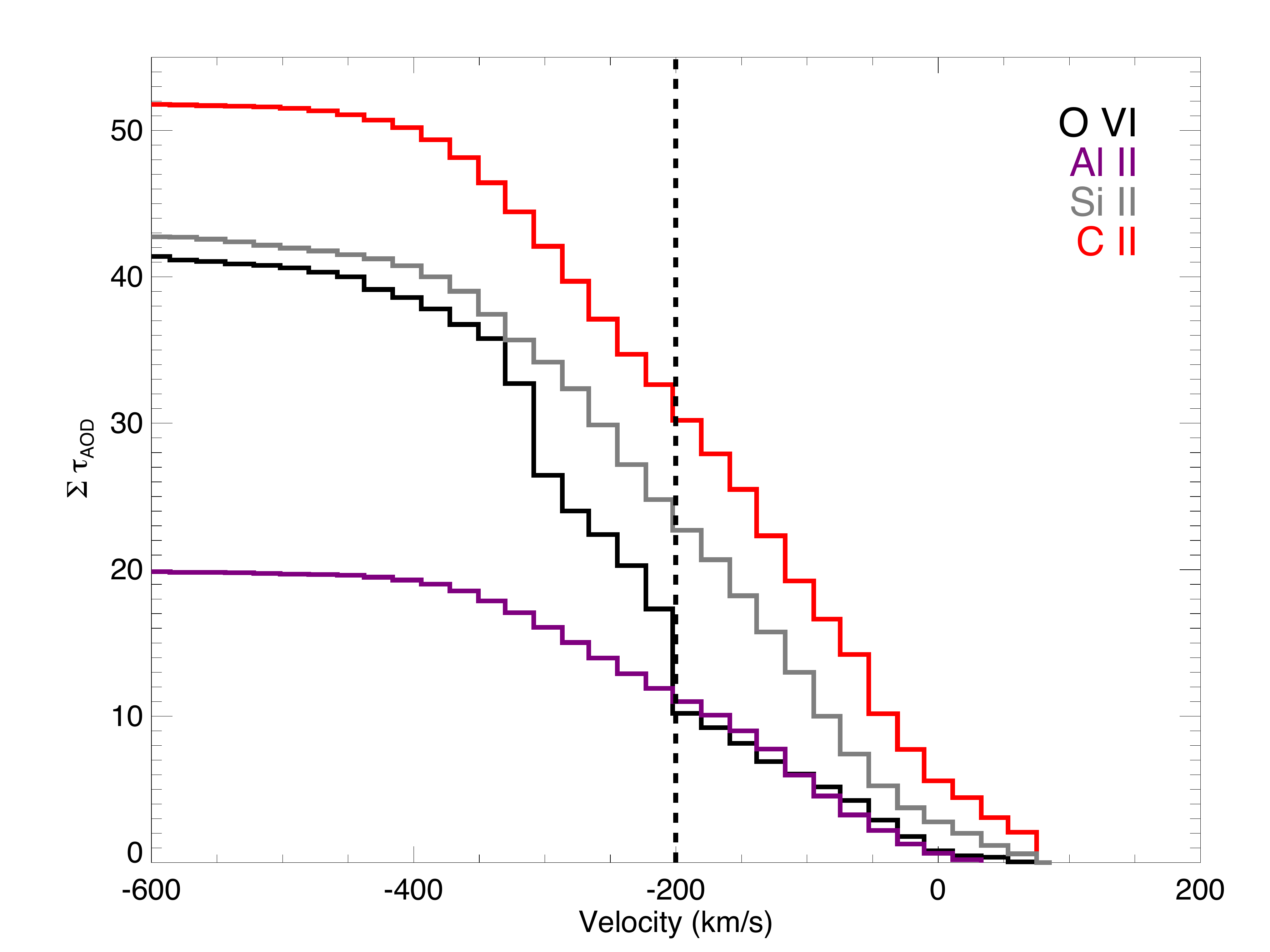}
\end{subfigure}
\caption{{\it Upper Panel:} Velocity-resolved apparent optical depth ($\tau_\mathrm{AOD}$) distributions for four transitions: \ovip~1032\AA\ (black line), \ciip~1335\AA\ (red line), \siiip~1260\AA\ (gray line) and \aliip~1670\AA\ (purple line). The \cii and \siii lines are saturated over the entire profile, while the \alii line is not. {\it Lower Panel:} Cumulative $\tau_\mathrm{AOD}$ distribution summed from positive velocities to negative velocities for the same ions as the upper panel. The \ovi $\Sigma\tau_\mathrm{AOD}$ matches the \alii $\Sigma\tau_\mathrm{AOD}$ from $0$~\kms to $-200$~\kms (marked by the dashed vertical line), while it matches the \siii and \cii distributions at velocities blueward of $-200$~\kmsp.}
\label{fig:op}
\end{figure}

\begin{figure}
\includegraphics[width = 0.5\textwidth]{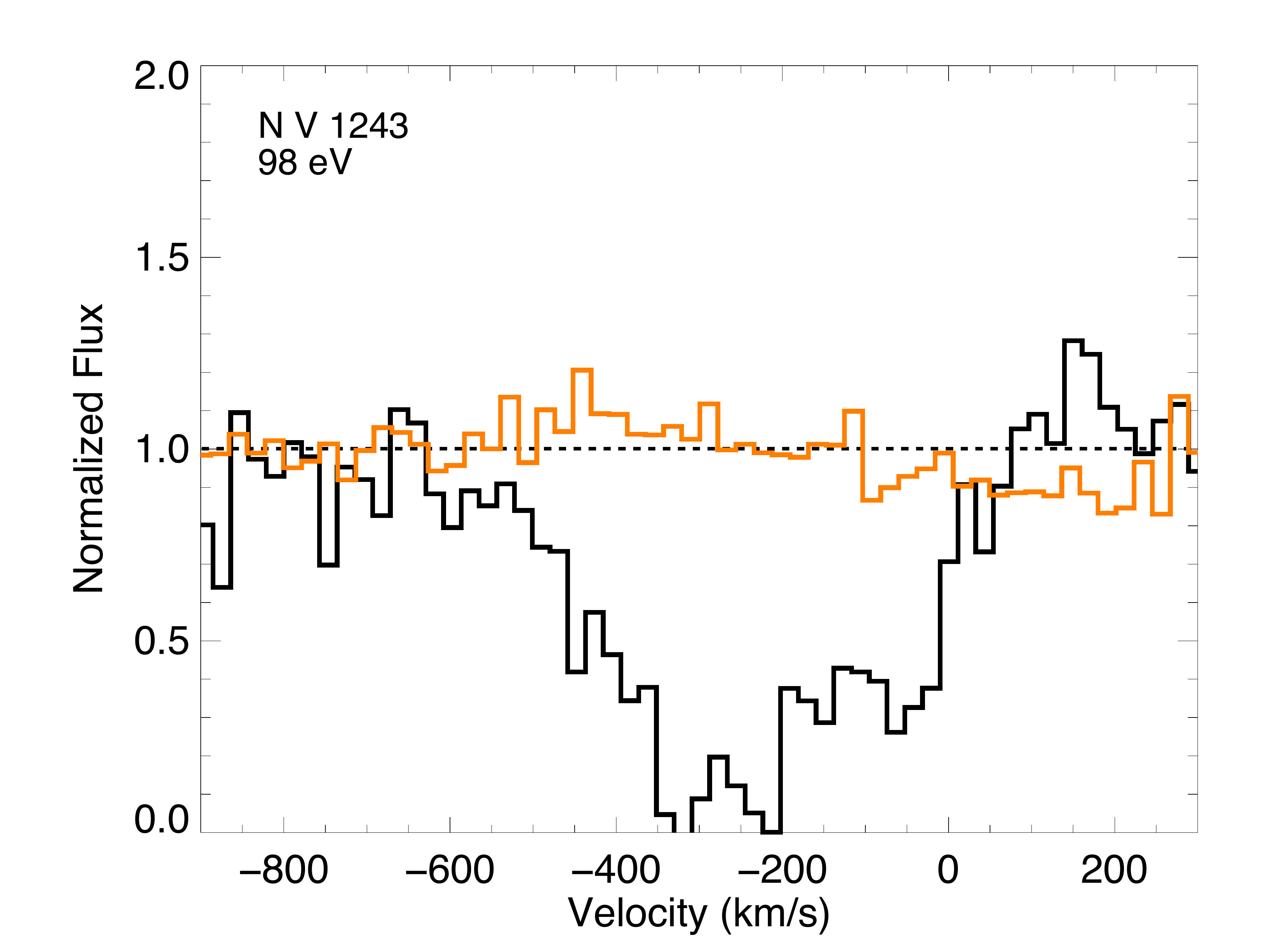}
\caption{Comparison of the stellar continuum-normalized absorption profiles for the \nvp~1243\AA\ transition (orange line) and the \ovip~1032\AA\ transition (black line). The dashed black line marks the unity flux level. The \nvp~1243\AA\ line is not detected in the spectrum, placing an upper limit on the \nv column density. }
\label{fig:nvlineprof}
\end{figure}

The stellar continuum-normalized \ovip~1032\AA\ line profile is shown in all panels of \autoref{fig:lineprof} as the black line, with low-ionization lines overplotted in colors. The \ovip~1032\AA\ line is broad (full width at half maximum of 150~\kmsp) and blueshifted relative to the stellar continuum by an average of -246~\kmsp, with deep absorption at blue velocities.

In \autoref{fig:lineprof} we show transitions of varying strength and ionization potential. The covering fraction and optical depth govern the line profiles, and we use the equivalent width ratios of doublet absorption lines to demonstrate whether a transition is optically thick. If the weaker line in a doublet is optically thin then the ratio of the two equivalent widths will be equal to the ratio of the two transition's $f\lambda$, where $f$ is the $f$-value of the line. Meanwhile, if the line is optical thick, the ratio of the two equivalent widths will be 1. The equivalent width ratio between \siiv 1393\AA\ and \siiv 1402\AA\ is 1.6, implying that the \siiv line is not strongly saturated because the $f\lambda$ ratio is 2.0. Similarly, in \autoref{fig:si2} we show the \siii 1808\AA, 1527\AA, and 1260\AA\ transitions which have $f$-values of 0.0025, 0.13, and 1.2, respectively. The \siiip~1260\AA\ transition is saturated over the entire profile, while the \siiip~1808\AA\ profile is only marginally detected (3$\sigma$). The \siiip~1527\AA\ line resides between these two extremes.  

Singlets, like \oip~1302\AA\ and \aliip~1670\AA, cannot have their saturation levels diagnosed, but \autoref{fig:lineprof} shows that these lines have similar profiles to \siiv 1393\AA. Meanwhile, lines like \ciip~1335\AA, \siiiip~1206\AA, and \siiip~1260\AA\ are nearly black at all velocities, implying that they are saturated at all velocities. Consequently, \autoref{fig:lineprof} shows four strong lines (\siiiip~1206\AA, \siiip~1260\AA, \ciip~1335\AA, and \civp~1548\AA) and four relatively weak lines (\oip~1302\AA, \siivp~1393\AA, \aliip~1670\AA, and \aliiip~1863\AA). 

By visual inspection of \autoref{fig:lineprof}, we compare the \ovi and low-ionization line profiles. The shape and depth of the velocity profile is insensitive to the ionization potential of the line: the \siiip~1260\AA\ profile has a lower ionization potential than the \siivp~1393\AA\ line, but the \siii profile extends to larger velocities.  At velocities between $+100$ and $-200$~\kmsp, the shape and depth of the \ovi profile resembles those of the weak lines like \oip~1302\AA, \aliip~1670\AA, \aliiip~1863\AA, and \siivp~1393\AA; while at velocities blueward of $-200$~\kmsp, the \ovi profile resembles the strong \siiip~1260\AA\ and \ciip~1335\AA\ lines. Consequently, the \ovi profile has two regimes: a low-velocity regime that follows the weak lines, and a high-velocity regime that follows the saturated lines. This indicates that the \ovi profile is sensitive to the optical depth at low-velocities, but it is saturated at high-velocities. We return to this in \autoref{profile}.

\subsection{Velocity-resolved apparent optical depth profiles}

The strength of a line is approximated by the optical depth ($\tau$), a proxy for the column density. The apparent optical depth \citep[$\tau_\mathrm{AOD}$;][]{savage} gives a lower limit of $\tau$ because the covering fraction ($C_f$), resolution effects, and saturation can increase the actual $\tau$ (see \autoref{profile}). In the upper panel of \autoref{fig:op}, we explore how the $\tau_\mathrm{AOD}$ of four ions evolves with velocity. The saturated \ciip~1334\AA\ and \siiip~1260\AA\ distributions are similar, while the \aliip~1670\AA\ distribution is a factor of 2.3 smaller. The \alii $\tau_\mathrm{AOD}$ does not rise much above 1, while \cii and \siii are above two for all velocities between 0 and $-350$~\kmsp, and decline collectively at bluer velocities.

The \ovip~1032\AA\ $\tau_\mathrm{AOD}$ follows the \alii $\tau_\mathrm{AOD}$ from 0 to $-200$~\kmsp, and rapidly increases at bluer velocities. At the highest velocities, the \ovi profile declines on the same trajectory as the \cii and \siii profiles. The velocity-resolved $\tau_\mathrm{AOD}$ follows the two-regime scenario for the \ovi profile outlined above.

The two-regime behavior is emphasized when we sum the $\tau_\mathrm{AOD}$ with velocity  (lower panel of \autoref{fig:op}). Again, the \ovi $\tau_\mathrm{AOD}$ rapidly diverges from the \alii profile at velocities blueward of $-200$~\kms (marked by the vertical dashed line), and approaches the \cii and \siii distributions. At $-200$~\kms the \ovi profile shifts from behaving as a weak line (like \aliip~1670\AA) to behaving as a strong line (like \ciip~1335\AA\ or \siiip~1260\AA). In \autoref{profile} we use the weak \siiv doublet to explore a possible physical mechanism for these two regimes.

The similarity between the \ovi and low-ionization line profiles indicates that the different ions are co-moving because they have similar absorption at similar velocities. Moreover, it appears that the \ovi transition strengthens at velocities blueward of $-200$~\kmsp, and the profile declines in tandem with the strongest transitions at the bluest velocities. Since these lines all have different optical depths, the shared decline at high velocities is likely due to a declining covering fraction. The covering fraction relates the physical sizes of the outflow, consequently, the similar covering fractions indicate that the \ovi is also co-spatial with low-ionization gas.

\subsection{Velocities of photoionized and transitional gas}
We measure the velocities for each transition in two ways (\autoref{tab:lines}). The first way is the central velocity (\vcenp), which is defined as the velocity at half of the total equivalent width of the line. \vcen is influenced by zero-velocity absorption \citep{chisholm15}, or by resonance emission lines \citep{prochaska2011, scarlata}, therefore we also measure the velocity at which the line profile reaches 90\% of the continuum on the blue-side of the profile (\vnp). We estimate the errors on each quantity by producing 1000 realizations of the data by multiplying each observed pixel by a random number drawn from a Gaussian distribution centered on zero with a standard deviation equal to the error on the flux.

The velocities of the individual transitions confirm what we saw by eye in \autoref{fig:lineprof}: strong transitions (\siii 1260\AA, \siiiip~1206\AA, and \ciip~1334\AA) have large \vnp, while weaker transitions (\aliip~1670 and \siiv 1402\AA) have significantly smaller \vnp. Since \ovi follows the weaker lines at low velocity and the stronger lines at high velocity, the \vcenp--which is weighted by the equivalent width--shifts bluewards. Therefore, \ovi has the largest \vcen of any transition, even though other transitions have larger (\siiip, \siiiip), or comparable (\ciip), \vnp. The two-regime behavior of \ovi may explain why previous studies, that only quantized the outflow velocity with \vcenp, found a larger \vcen \text{for \ovi} than the low-ionization lines, even though the \ovi absorption profiles did not extend bluer than the low-ionization lines \citep[Figure 3 of][]{heckman01, grimes2009}.

\subsection{The non-detection of \ion{N}{V}}
\label{nv}

Before preceding, one notable exception to these observations is the \nvp~1243\AA\ line (see \autoref{fig:nvlineprof}). Strangely, interstellar \nvp~1243\AA\ is not detected in absorption in the spectrum of J1226+2152, even though \nv has an ionization potential between the observed \ovi and \civ transitions. \nvp~1243\AA\ is a relatively strong transition ($f$-value of 0.08) arising from a cosmologically abundant element \citep[N$/\mathrm{H} = 7.5\times10^{-5}$;][]{jenkins}, indicating that there is negligible gas in the \nv ionization state. \nv is also rarely detected in lower redshift galactic outflows \citep{chisholm16}, providing  a clue about the ionization structure of galactic outflows. We will use the \nv non-detection in \autoref{o6} when we explore how \ovi is produced.

\section{MODELING THE ABSORPTION LINE PROFILES}
\label{profile}
\begin{figure*}
\includegraphics[width = \textwidth]{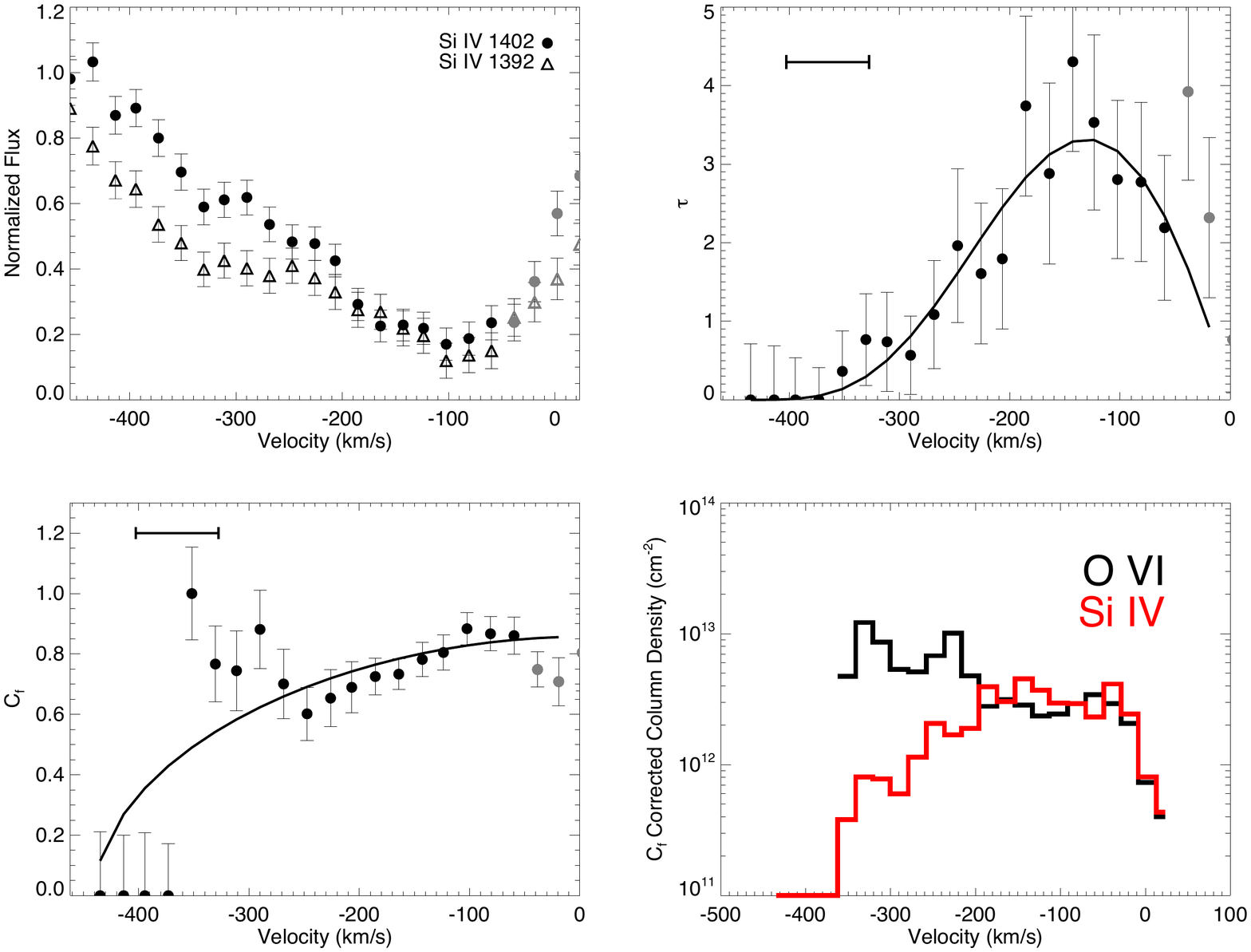}
\caption{\siiv profile fits. The upper left panel shows the \siiv 1402\AA\ (filled circles) and \siiv 1393\AA\ (triangles) profiles. The grey points are excluded from the fit due to contamination with resonance emission, as determined from the \siiip$^\ast$ emission line. These profiles are used to calculate the velocity-resolved optical depth ($\tau$; upper right panel) and covering fraction ($C_f$; lower left panel) using the radiative transfer equation (\autoref{eq:cfobs}). We simultaneously fit a model to the $C_f$ and $\tau$ distributions (\autoref{eq:cfbeta}) as shown by the solid lines in the two panels. In the lower right panel we use the observed \siiv $C_f$ to correct the \ovi (black) and \siiv (red) column densities for partial coverage, assuming that \ovi and \siiv have similar $C_f$. At velocities blueward of -200~\kms the column density profiles diverge: the \siiv column density rapidly decreases while the \ovi column density increases. The velocity resolution of 75~\kms is indicated by bars in the $\tau$ and $C_f$ panels.}
\label{fig:proffits}
\end{figure*}
In the previous section we found that the \ovi line profile has two regimes bisected at $-200$~\kmsp: the depth of the \ovi line matches the weak lines at redder velocities, while the depth matches the strong lines at bluer velocities. Here we model the line profiles to explore the origin of this two-regime behavior.

\subsection{Velocity-resolved optical depths and covering fractions}

There are two degenerate ways to change the depth of an absorption profile: $C_f$ and $\tau$. Both these quantities change with velocity to produce the observed profile. $C_f$ is the fraction of the total area of the background continuum source that the absorbing gas covers. A $C_f$ of 1 means that the gas completely covers the background stellar continuum, while a $C_f$ of 0 means that the gas does not cover the background source. If $\tau$ is large, then the depth of the line is $1-C_f$.

The second way to change the depth of an absorption profile is through the optical depth, or approximately the column density ($N$). At $\tau$ greater than 3 the line becomes optically thick and $C_f$ completely sets the depth of the line. Meanwhile at lower optical depths some light passes through the gas, and the flux is no longer zero. Consequently, the depth of an absorption line is degenerate with both $\tau$ and $C_f$. The radiative transfer equation describes this degeneracy as:
\begin{equation}
F(v) = 1-C_f(v)+C_f(v) e^{-\tau(v)}
\label{eq:radtran}
\end{equation}
where $F(v)$ is the continuum normalized flux transmitted through the gas at a particular velocity ($v$). The only way for $F(v)$ to approach zero is if $C_f(v)$ approaches 1 and $\tau(v) > 3$ ($e^{-3} = 0.05$), otherwise the line has residual flux. This is seen in \autoref{fig:lineprof} and \autoref{fig:si2} where the weak lines (\oip~1302\AA, \siiip~1527\AA, \aliip~1670\AA, and \siivp~1393\AA) all have non-zero flux across their line profiles, while the strong lines (\siiip~1260\AA, \ciip~1335\AA) have an average depth of 0.1 between $0$ and $-300$~\kmsp, implying that $C_f(v)\approx0.9$. 

The degeneracy between $C_f$ and $\tau$ is broken with a doublet from the same ionic species. With a doublet, \autoref{eq:radtran} becomes a system of two equations where each line has the same $C_f$ and their $\tau$ is related by the ratio of their $f$-values. The \siivp~1402\AA\ and \siivp~1393\AA\ doublet has a $f$-value ratio of 2. The system of equations exactly solves for $C_f$ and $\tau$ in terms of the flux of the two transitions as
\begin{equation}
 \begin{aligned} 
    C_f(v) &= \frac{F_\mathrm{W}(v)^2-2F_\mathrm{W}(v) + 1}{\mathrm{F}_\mathrm{S}(v)-2\mathrm{F}_\mathrm{W}(v)+1}\\
    \tau(v) &= \ln\left(\frac{C_f(v)}{C_f(v)+F_\mathrm{W}(v)-1}\right)
\end{aligned}
\label{eq:cfobs}
\end{equation}
Where $F_W(v)$ is the flux of the weaker doublet line at a given velocity and $F_S(v)$ is the flux of the stronger line. We calculate the errors on $C_f(v)$ and $\tau(v)$ by preforming a Monte Carlo analysis at each velocity, similar to how we calculated the velocity errors. 

Using \autoref{eq:cfobs}, we solve for the \siiv $C_f$ and $\tau$ at each velocity (\autoref{fig:proffits}). At very low velocities the $\tau$ and $C_f$ are impacted by zero-velocity absorption and resonance emission \citep{prochaska2011, scarlata}. To assess the contamination from the emission, we use the \siiifsp~1197\AA\ non-resonant emission line to determine which velocities are contaminated by the emission. The \siiifs line is narrow, with emission only within $\pm50$~\kmsp, therefore, we exclude those velocities from consideration (grey points in \autoref{fig:proffits}). 

In the lower left panel of \autoref{fig:proffits}, $C_f(v)$ declines at velocities blueward of $-40$~\kms until velocities near $-300$~\kms where it increases from $\approx70$\% covered to fully covered, and then rapidly settles to zero coverage. The increase blueward of $-300$~\kms in $C_f$ is observed in the \siiip~1260\AA, \siivp~1393\AA, and \aliip~1670\AA\ profiles as a \lq{}\lq{}second component\rq{}\rq{} (\autoref{fig:lineprof}).

The \siiv $\tau(v)$ peaks at velocities near $-175$~\kmsp, and declines at velocities blueward of $-200$~\kms (upper right panel of \autoref{fig:proffits}). At the bluest velocities, the \siiv 1402\AA\ line profile is set by both $C_f(v)$ and $\tau(v)$ because $\tau(v) < 3$. Meanwhile, stronger lines, like \ciip~1335\AA\ and \siiip~1260\AA, still have $\tau(v) > 3$ at these velocities, and their line profiles do not change appreciably until velocities blueward of $-350$~\kms when $C_f(v)$ rapidly drops. The degeneracy between $\tau(v)$ and $C_f(v)$ dominates the line profiles of weak lines, while $C_f(v)$ determines the line profiles of strong lines.

\subsubsection{An origin of the two-regime behavior of \ovi}
\label{o6orgin}
\autoref{fig:proffits} illustrates why the \ovi profile has two regimes. At low-velocities the \ovi profile follows the $\tau$ dominated regimes of the weak lines, while at velocities blueward of $-200$~\kms \ovi is optically thick and the line profile is determined by $C_f$. This split at $-200$~\kms is the same velocity where the \siiv $\tau$ begins to decline, indicating that the \ovi $\tau$ increases as the \siiv $\tau$ decreases. 

To illustrate this, we take the observed \siiv $C_f$ from the lower left panel of \autoref{fig:proffits} and correct the \ovi column density for the partial coverage (bottom right panel of \autoref{fig:proffits}; see \autoref{colden} for details on this correction). At redder velocities the \siiv and \ovi column densities follow similar trends, but at bluer velocities the \siiv column density decreases as the \ovi column density increases. The increased \ovi column density means that \ovi becomes a strong line, like \siiip~1260\AA\ or \ciip~1335\AA, and the blue-wing of the \ovi profile is now fixed by $C_f$. This also means that at bluer velocities the \ovi line likely saturates, and the derived velocity-resolved column densities are lower limits. The transition of \ovi from an optically thin line to an optically thick line naturally explains the two-regime \ovi profile: at redder velocities the \ovi profile matches the weak transitions whose profiles are governed by changes in $\tau$ at a nearly constant $C_f$, while at bluer velocities the \ovi profile resembles the strong transitions whose profiles change as $C_f$ changes. Similar behavior is also observed in other high signal-to-noise ratio spectra in the \megasaura\ sample (see J1527+0652 in \autoref{fig:megasaura}).  

\subsection{A model describing the \siiv profile}
\label{proffits}
Above, we found that the \ovi column density rises as the \siiv column density declines. This suggests a relationship between the two phases of gas that we discuss further in \autoref{model}. To gain insight into this connection, here we explore a physical model for the \siiv profile. This model was fully introduced in \citet{chisholm16b}, and we refer the reader there for details on its derivation.
\begin{table*}
\caption{Parameter estimates to the simultaneous $\tau$ and $C_f$ fit of \autoref{eq:cfbeta}. Column 2 gives the optical depth normalization ($\tau_0$), Column 3 gives the maximum covering fraction ($C_f(R_\mathrm{i})$), Column 4 gives the radial velocity-law exponent ($\beta$; \autoref{eq:beta}), Column 5 gives the density power-law exponent ($\alpha$; \autoref{eq:den}), and Column 6 gives the covering fraction exponent ($\gamma$; \autoref{eq:cf}).  The first row gives the estimates for the high redshift galaxy J1226+2152, while the second row shows the parameters for the local star-forming galaxy NGC~6090 \citep{chisholm16b}. }
\begin{tabular}{cccccc}
\hline
Galaxy & $\tau_0$ & $C_f$($R_\mathrm{i}$) & $\beta$ & $\alpha$ & $\gamma$ \\
 \hline
J1226+2152 & $8.1 \pm 2.5$ & $0.86 \pm 0.03$ & $0.53 \pm 0.10$ & $-6.3 \pm 1.7$ & $-0.53 \pm 0.13$ \\
NGC~6090 & $4.8 \pm 1.4$ & $1.0 \pm 0.04$ & $0.43 \pm 0.07$ & $-5.7 \pm 1.5$ & $-0.82 \pm 0.23$ \\
\end{tabular}
\label{tab:lineprof}
\end{table*}

Physically, $C_f$ varies with velocity because the absorbing clouds cover a changing amount of the total area of the background stellar continuum. This change is caused by geometric dilution--the clouds occupy a smaller fraction of the total area at larger radii--or because the size of the outflowing clouds change with velocity \citep{martin09, steidel10}. A simple geometric model relates changes in $C_f$ to changes in the area occupied by the absorbing gas as a power-law with
\begin{equation}
C_f (r) = C_f(R_\mathrm{i}) \left(\frac{r}{R_\mathrm{i}}\right)^\gamma
\end{equation}
where $C_f$(R$_\mathrm{i}$) is the covering fraction at the initial radius (R$_\mathrm{i}$) and $\gamma$ is an unknown power-law exponent. Introducing normalized coordinates simplifies this relation, such that
\begin{equation}
C_f(x) = C_f (R_\mathrm{i}) x^\gamma
\label{eq:cf}
\end{equation}
where $x = r/R_\mathrm{i}$. If the outflowing clouds remain the same size, $C_f$ decreases with radius as a power of $\gamma = -2$ as the projected area increases as r$^2$. For a local star-forming galaxy, NGC~6090, \citet{chisholm16b} find that the $C_f$ evolves as r$^{-0.8}$, less rapidly (lower $\gamma$) than expected from geometric dilution, but consistent with the outflow adiabatically expanding as it accelerates.

Changes in $\tau(v)$ come from two sources: changes in the density of the absorbers ($n(v)$) and the velocity distribution of the absorbers. A power-law describes the radial change in density, such that
\begin{equation}
n\left(r\right) = n_0 \left(\frac{r}{R_\mathrm{i}}\right)^\alpha= n_0 x^\alpha
\label{eq:den}
\end{equation}
where $\alpha = -2$ if the outflow is a mass-conserving spherical outflow, and $\alpha < -2$ if gas is removed from the outflow, or if the geometry significantly deviates from a spherical outflow \citep{fielding}. Possible ways to remove gas from the outflow are ionizing the outflow to a different ionization state \citep{chisholm16b} or if gas exits the outflow as a galactic fountain \citep{leroy15}. Again for NGC~6090, \citet{chisholm16b} find $\alpha = -5.7$, possibly because a hot outflow destroys the low-ionization gas and incorporates it into a hotter outflow. 

Changing how the gas is distributed in velocity also changes $\tau(v)$. Often times Gaussian or Lorentzian velocity distributions are assumed, where the absorbers are spread out in velocity space due to their thermal (or turbulent) motions \citep{rupkee2005, martin09, chen10}. However, these assumed velocity distributions do not describe galactic outflows in which the radial velocity gradient broadens the observed line profile. The outflow velocity gradient measures how rapidly the velocity changes with radius ($dv/dr$). If $dv/dr$ is large, then the velocity of the outflow rapidly changes while only traveling a short radial distance. In this case, there are fewer total absorbers per velocity interval, and correspondingly the optical depth is lower. Conversely, a small $dv/dr$ means that the outflow accelerates gradually with radius, piling more absorbers into each velocity interval. The Sobolev approximation \citep{sobolev} defines $\tau$ in terms of the density and radial velocity gradient as
\begin{equation}
    \tau(v) = \frac{\uppi e^2}{\mathrm{mc}} f \lambda_\mathrm{0} n(v) \frac{dr}{dv}
    \label{tau}
\end{equation}
where $f$ is the $f$-value of the transition, $\lambda_\mathrm{0}$ is the transition's rest wavelength, and m is the mass of the electron.

The velocity gradient is not known a priori, but we fit for it from the shape of $\tau(v)$ and $C_f(v)$ (\autoref{fig:proffits}). Analytic relations for the acceleration (or deceleration) of outflows find that the velocity changes with radius \citep{murray05} as a $\beta$-velocity law \citep{lamers}, such that
\begin{equation}
v = v_\infty (1-\frac{R_\mathrm{i}}{r})^{\beta} =  v_\infty (1-\frac{1}{x})^\beta
\label{eq:beta}
\end{equation}
where v$_\infty$ is the maximum observed velocity (440~\kms for \siiv 1402\AA). It is important to note that \autoref{eq:beta} does {\it not} require that the outflow accelerates: a negative $\beta$ produces a decelerating profile. Rather, we fit for $\beta$ from the \siivp~1402\AA\ profile, while allowing for $\beta$ values between $-5$ and $+5$. The shape of $\tau(v)$ and $C_f(v)$ define whether the outflow is accelerating or decelerating. In \citet{chisholm16b}, NGC~6090 has $\beta = 0.43$, which is consistent with a r$^{-2}$ force law opposed by gravity, such as ram pressure \citep{chisholm16b}. 

The derivative of \autoref{eq:beta} defines the velocity gradient, allowing for the $C_f$ (\autoref{eq:cf}) and $\tau$ (\autoref{tau}) distributions to be rewritten in terms of the normalized velocity ($w = v/v_\infty$) as
\begin{equation}
\begin{aligned}
\tau\left(w\right) &= \frac{\uppi e^2}{mc} f \lambda_\mathrm{0} \frac{R_\mathrm{i}}{v_\infty} n_\mathrm{4, 0} x^\alpha \frac{dx}{dw}= \tau_0 \frac{w^{1/\beta-1}}{\beta(1-w^{1/\beta})^{2+\alpha}}\\
C_f(w) &=  \frac{C_f (R_i)}{\left(1-w^{1/\beta}\right)^\gamma}
\label{eq:cfbeta}
\end{aligned}
\end{equation}
where n$_\mathrm{4,0}$ is the \siiv density at $R_\mathrm{i}$. The constant preceding the $\tau$ distribution is rewritten as
\begin{equation}
\tau_0 = \frac{\uppi e^2}{\mathrm{mc}} f \lambda_\mathrm{0} \frac{\mathrm{R}_\mathrm{i}}{\mathrm{v}_\infty} \mathrm{n}_\mathrm{H,0} \chi_\mathrm{Si 4} \mathrm{(Si/H)}
\label{eq:tau0}
\end{equation}
Where n$_\mathrm{H,0}$ is the total hydrogen density at the base of the outflow,  $\chi_\mathrm{Si 4}$ is the \siiv ionization fraction, and Si/H is the silicon to hydrogen abundance. With the ionization models of \autoref{cloudy}, we solve for $R_\mathrm{i}$ in \autoref{eq:tau0} to derive the initial radius of the outflow.

We use {\small MPFIT} \citep{mpfit} to fit for the five parameters in \autoref{eq:cfbeta} ($\tau_0$, $C_f(R_\mathrm{i})$, $\beta$, $\alpha$, $\gamma$; see \autoref{tab:lineprof}). In \autoref{tab:lineprof} we compare the fitted parameters to the local star-forming galaxy NGC~6090 \citep{chisholm16b}. $\beta$, $\alpha$ and $\gamma$ describe the radial variation of $C_f$ (\autoref{eq:cf}),  density (\autoref{eq:den}), and velocity (\autoref{eq:beta});  these exponents are similar to local galaxies. This implies that the outflow properties of high and low redshift galaxies vary similarly with radius \citep{rigbyb}. Intriguingly, the $C_f(R_\mathrm{i})$ of 0.86 is roughly consistent with the maximum depth of the \siiip~1260\AA\ and \ciip~1335\AA\ lines being near 0.1 in normalized flux units, further emphasizing that these lines are optically thick and dominated by $C_f$.

These models illustrate many interesting features of the \siiv profiles. For instance, the rapid acceleration ($\beta = 0.53$) of the outflow means that the outflow initially does not travel a large distance in each velocity interval. This rapid acceleration means that the dynamical time, as defined by the velocity profile (\autoref{eq:beta}), is only $\sim0.2$~Myr. Since the density decreases with radius, not velocity, the rapid acceleration ensures that the density does not decrease appreciably in each redder velocity interval. In fact, \autoref{eq:den} implies that at -200~\kms the density is still 20\% of the initial density. This allows the \siiv and \alii lines to remain strongly detected, but the weaker \siiip~1808\AA\ transition is undetected blueward of $-200$~\kms (\autoref{fig:si2}). By the time the outflow reaches -300~\kms the \siiv density is now 1.5\% of the original density, and at -400~\kms the density is $1\times 10^{-5}$ times the original density. At these densities the \siiv optical depth drops below the detection threshold, and the line is no longer observed. Meanwhile, the \siiip~1260\AA\ line has an $f$-value 5 times larger and an ionization correction that is twice that of \siivp~1402\AA\ (see \autoref{cloudy}), implying that the \siii $\tau$ is roughly 10 times larger than the \siiv $\tau$. From \autoref{fig:proffits}, the \siiip~1260\AA\ line is still optically thick at $-300$~\kmsp, with $\tau$ near 10. Even at large velocities, \siiip~1260\AA\ is optically thick and the profile is determined by $C_f$. 

Finally, \autoref{eq:cf} explains why $C_f$ evolves slowly at redder velocities. The rapid acceleration means that the outflow does not travel far in these velocity intervals. At bluer velocities the outflow covers more distance ($dv/dr$ is smaller), decreasing the $C_f$ and shaping the line profiles of the strong lines (\siii~1260\AA, \ciip~1335\AA, and \ovip~1032\AA). 

At the moderate spectral resolution of the \megasaura\ data, it is important to check that the resolution does not dramatically affect the measured $C_f$. To do this, we make a synthetic line profile using the fitted properties from \autoref{tab:lineprof}, and then convolve the profile to the 75~\kms resolution of the \megasaura\ data. We then measure the deepest portion of the profile and find that it increases by 0.09 normalized flux units. This increase is within the median $C_f$ error of 0.13 normalized flux units  between $-400$ and 0~\kms (see \autoref{fig:proffits}), therefore we conclude that the impact of the resolution is within our quoted $C_f$ errors. 

\section{PHOTOIONIZATION MODELING}
\label{cloudy models}

In the previous section, we used the line profiles to show that the \ovi column density increases as the \siiv density declines, even though the two phases are co-spatial. This suggests that the \ovi column density is decoupled from the \siiv column density, and the low-ionization gas is created by a different mechanism than the transitional gas. In \citet{chisholm16}, the low-ionization equivalent widths are consistent with predictions from {\small CLOUDY} photoionization models \citep{ferland}, if the observed stellar continuum is the ionizing source. Here, we use the observed column densities and the stellar continuum fit to predict the column densities of the individual transitions. This tests whether the \ovi is also photoionized, or if separate ionization mechanisms produce the observed \ovi and photoionized gas.

\subsection{Measuring observed column densities}
\label{colden}
The observed column density ($N$) of a given transition is the product of the outflow's metallicity, density, and ionization structure. To characterize the ionization structure, we measure the integrated column density of each transition using a modified apparent optical depth method \citep{savage}  which accounts for non-unity covering of the source (see \autoref{proffits}) as 
\begin{equation}
    N = \frac{3.77 \times 10^{14}~\text{cm}^{-2}}{\lambda_0[\text{\AA}] f} \int \ln{\left( 
    \frac{C_f(R_\mathrm{i})}{C_f(R_\mathrm{i}) + F_\mathrm{o}(v) - 1} \right)} \mathrm{d}v
\label{eq:colden}
\end{equation}
where $\lambda_0$ is the rest wavelength of each transition, $f$ is the oscillator strength of each transition, and $F_\mathrm{o}$($v$) is the continuum-normalized flux. We use the $C_f(R_\mathrm{i}$) from the \siivp~1402\AA\ transition because many of these lines are singlets, and we cannot derive their velocity-resolved $C_f$ distributions. Additionally, \citet{chisholm16} show that $C_f$ does not vary appreciably from transition to transition. This can be seen in \autoref{fig:lineprof} where $C_f$ appears to be near 0.9 for the strong transitions (where $C_f = 1-F$ for saturated lines).  The errors are calculated similar to the velocity errors, and a 10\% uncertainty is included, in quadrature, to account for the continuum normalization uncertainty. The values are given in \autoref{tab:lines}.  

Many of these transitions could be heavily saturated. \aliiip~1862\AA\ and \siivp~1402\AA\ are doublets, and their equivalent width ratios diagnose saturation (discussed in \autoref{lineprof}). The observed \aliii and \siiv doublet ratios are 1.4 and 1.6, while their $f$-value ratios are both 2. This implies that these transitions are not heavily saturated.

The wavelength coverage and sensitivity of the \megasaura\ data also allows us to measure lines that are not typically measured \citep[see][for another example]{pettini2002}. The \siiip~1808\AA\ line has an $f$-value of 0.0025, and is unsaturated. At a signal-to-noise ratio near 10, we marginally detect this line above the continuum noise (a 3$\sigma$ detection; see \autoref{fig:si2}). The logarithm of the \siiip~1808\AA\ column density is $15.57  \pm 0.14$~cm$^{-2}$, which is consistent, within 2$\sigma$, with the column density measured from \siiip~1304\AA\ and \siiip~1526\AA\ (see \autoref{tab:lines}). The photoionization models of the next section rely on the uncertainties of the column densities, and large statistical uncertainties, like the errors we measure for \siiip~1808\AA, do not constrain the models. Therefore, we use the \siiip~1526\AA\ column density (as denoted by the \lq{}{}\lq{}c\rq{}{}\rq{}{} in \autoref{tab:lines}), but note that the \siii lines could be marginal saturated.

\subsection{Can photoionization models reproduce the \ovip?}
\label{cloudy}
\begin{figure}
\includegraphics[width = 0.5\textwidth]{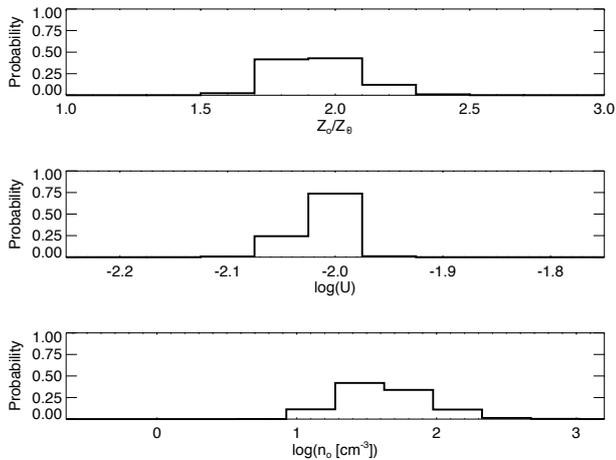}
\caption{Probability density functions for the three parameters used to generate the {\sc CLOUDY} models: outflow metallicity (Z$_\mathrm{o}$), ionization parameter (U), and hydrogen density at the base of the outflow (n$_\mathrm{H,0}$). The expectation values for Z$_\mathrm{o}$, log(U), and n$_\mathrm{H,0}$ are 1.94~Z$_\odot$, -2.01~dex, and 42~cm$^{-3}$.}
\label{fig:pdfs}
\end{figure}

\begin{table}
\caption{{\small CLOUDY} parameter grid used to estimate the ionization parameter (log(U)), hydrogen density at the base of the outflow (n$_\mathrm{H,0}$) and metallicity of the outflow ($Z_\mathrm{o}$). The second column gives the parameter range used in the Bayesian analysis and the third column gives the number of bins for each parameter. The expectation values and standard deviations of the parameters for J1226+2152 are given in the fourth column, and the fifth column shows the estimates for NGC~6090, a local star-forming galaxy.}
\begin{tabular}{lcccc}
\hline
Parameter & Range & Bins & Expected Value & NGC~6090  \\
(1) & (2) & (3)& (4) & (5)\\
 \hline
log(U) & $-1.5 \rightarrow -2.5$~dex & 21 & $-2.01 \pm 0.02$ & -1.85 \\
log(n$_\mathrm{H,0}$) & $-0.65 \rightarrow 3.2$~dex & 12 &  $42 \pm 31$~cm$^{-3}$ & 19~cm$^{-3}$ \\
$Z_\mathrm{o}$ & $0.1 \rightarrow 3.0$~Z$_\odot$ & 16 & $1.94 \pm 0.15$~Z$_\odot$ & 1.6~Z$_\odot$ \
\end{tabular}
\label{tab:cloudy}
\end{table}

The ionization structure depends on the input energy that heats the gas as well as how efficiently the gas loses energy by radiatively cooling. This cooling depends on the density, temperature, and metallicity of the gas. Consequently, the observed ionization structure sensitively depends on the ionizing spectrum, the density of the outflow, and the metallicity of the outflow. 

We characterize the O and B stars ionizing the outflow with the measured {\small STARBURST99} stellar continuum from the wavelength regime redward of 1240\AA. We then model the ionization structure of the outflows using {\small CLOUDY}, v13.03 \citep{ferland}, models that use these O and B stars as the ionizing source. In {\small CLOUDY}, we use an expanding spherical geometry, with the density profile determined in \autoref{proffits}, and the initial covering fraction ($C_f(R_\mathrm{i})$). By including the fitted density profile from \autoref{proffits}, we incorporate the decreasing outflow density with increasing velocity, which creates the two-regime \ovi behavior (\autoref{o6orgin}). We use {\small CLOUDY}'s \ion{H}{ii} abundances, with the Orion nebular dust grain distribution \citep{baldwin1991}, to account for dust scattering and depletion of metals onto grains. We do not change the relative abundances (like $\alpha$-enhancement, or Fe-enrichment), which may have second order effects. We then scale the abundances by constant factors to enrich or deplete the gas according to \autoref{tab:cloudy}. We stop running {\small CLOUDY} once the temperature has dropped to 500~K, and include a cosmic ray background \citep{indriolo}. 

From a given set of ionization parameters (U), outflow metallicities ($Z_\mathrm{o}$), and hydrogen densities at the base of the outflow (n$_\mathrm{H,0}$); {\small CLOUDY} estimates the total integrated column densities of the various ions along the line-of-sight, which we compare to the observed column densities using Bayesian inference \citep{kauffmann2003, brinchmann2004}. We use the large grid of {\small CLOUDY} models from \autoref{tab:cloudy} to calculate the probability of each model, given the observed column densities, using the likelihood function of
\begin{equation}
L \sim \exp\left(-\chi^2\right)
\end{equation}
where $\chi^2$ is the chi-squared function of the observed \oip~1302\AA, \siii~1527\AA, \aliip~1670\AA, \aliiip~1862\AA, and \siiv 1402\AA\ column densities. Note that the observed \ovi is not included in these photoionization models because we are trying to predict the \ovi column density, given the observed low-ionization column densities, to determine if the observed \ovi can be produced by photoionization. We then marginalize over nuisance parameters to produce the probability density functions (PDFs) for each parameter (\autoref{fig:pdfs}). We take the expectation values and standard deviations of each of these PDFs as the parameter estimates of log(U), Z$_\mathrm{o}$ and n$_\mathrm{H,0}$ (\autoref{tab:cloudy}). 

We then produce a {\small CLOUDY} model using the expectation values from the Bayesian inference. This gives the abundances and ionization fractions of each ion. For instance, the {\small CLOUDY} model predicts that the neutral fraction of the outflow is 3\%, the \siiv ionization fraction is 14\%, and the gas phase Si/H abundance is 10$^{-5.11\pm0.02}$. The errors on the abundances are calculated by creating {\small CLOUDY} models using the standard deviations of the individual parameters. Additionally, the model predicts that the dominant silicon ionization state is \siiii (containing 55\% of the total silicon), and the outflow has a total hydrogen column density of 10$^{20.75\pm0.04}$~cm$^{-2}$. With these photoionization models, and the fitted $\tau_0$ value, \autoref{eq:tau0} solves for the initial radius of the outflow from the ionizing source of $R_\mathrm{i}~=~28$~pc. This result is marginally smaller than the 33-101~pc range found for local galaxies in \citet{chisholm17}.

J1226+2152 has a stellar mass upper-limit  of $<10^{9.5}$~M$_\odot$ \citep{wuyts}, and the derived Z$_\mathrm{o}$ of $1.94  \pm 0.15$~Z$_\odot$ is significantly larger than the expected ISM metallicity from the mass-metallicity relation. Local low-mass ($\sim10^7$~M$_\odot$) dwarf galaxies have similarly enriched galactic outflows, relative to their ISM metallicities (Chisholm et al. in preparation), which is proposed to drive the steep-portion of the observed stellar mass-metallicity relationship because outflows remove a substantial fraction of the metals produced during star formation \citep{tremonti04}. Even if the photoionization modeling overestimates Z$_\mathrm{o}$, the relative results presented below do not change because we adopt the same metallicity for both the photoionized and \ovi phases. Further, decreasing Z$_\mathrm{o}$ does not allow for the photoionization models to reproduce the observed \ovi column densities. 

The fitted {\small CLOUDY} column densities have a median deviation of 0.2~dex from the observed values (see the last column of \autoref{tab:lines}). The observed \siiv column densities, as well as the lower limits of the blended \civ lines, are consistent with the {\small CLOUDY} model, while \ovi is not produced at all (see \autoref{tab:lines}). Given the observed stellar continuum and the observed low ion column densities, the {\small CLOUDY} models do not produce nearly enough \ovi to match the observations. Another ionization mechanism must produce the observed \ovip.

\section{DISCUSSION}
\subsection{How is the \ion{O}{iv} created?}
\label{o6}
Since the photoionization models are unable to create nearly enough \ovip, we explore possible alternative mechanisms to produce this transitional gas. The observations provide four constraints:

\begin{enumerate}
    \item \ovi has a large integrated column density of 10$^{15.35}~\mathrm{cm}^{-2}$ with velocity-resolved \ovi densities near 2-10~$\times$~10$^{12}$~cm$^{-2}$~km$^{-1}$~s (\autoref{fig:proffits}).
    \item \nv is not detected, with upper limits of the \nv to \ovi ratio of log($N$(\nvp)/$N$(\ovip))~<~-1.2 (see \autoref{nv} and \autoref{fig:nvlineprof}). 
    \item \civ and \siiv column densities are consistent with the column densities from the photoionization models, implying that the mechanism that creates the \ovi produces negligible \siiv and \civp.
    \item Low-ionization lines (like \oip, \siiip, \aliip, etc.) have column densities consistent with the photoionization models. 
\end{enumerate}
Below we explore two excitation mechanisms that satisfy these four constraints. These two mechanisms are not an exhaustive list of ways to create \ovi \citep[see ][ for reviews]{spitzer96, indebetouw, wakker12}, but the above constraints eliminate many other mechanisms. Turbulent mixing layers between hot and cool gas, for example, require a log($N$(\nvp)/$N$(\ovi)) near -0.5 \citep{slavin93, kwak}, which is inconsistent with constraint (ii) above. Additionally, collisional ionization equilibrium \citep{sutherland93}, galactic fountain models \citep[]{shapiro91}, and shock models \citep{allen} do not satisfy constraint (iv) \citep{chisholm16}.

Here, we explore two mechanisms that do satisfy the four criteria above. The two models explain opposite phenomena: cool gas evaporating into a hot medium (\autoref{conduction}) and cool gas condensing out of a hot medium (\autoref{condense}). In the next two subsections we explain the two mechanisms and explore their implications for the ionization fractions of the transitional outflow. We stress that both models have many unconstrained parameters, and rather than trying to derive estimates of the total transitional gas, we maximize the values to predict upper limits. These models must be considered order of magnitude estimates, rather than precise calculations.  

\subsubsection{Conductive evaporation}
\label{conduction}
Thermal conduction  transfers thermal energy from a hot ($>10^7$~K) wind to cooler gas, heating and evaporating the cool gas \citep{mckeecowie}. As the interaction time increases, conduction heats more of the cool gas. This decreases the total amount of cool gas, while increasing the amount of hot and transitional gas \citep{indebetouw}. In this case, the \ovi is short-lived as the hot outflow further heats the transitional gas to higher temperatures, and the \ovi is only observed during this brief transition period.

Many studies explore the evaporation of cool gas by conduction with different assumptions for the direction of the hot flow and the orientation of the magnetic fields \citep{mckeecowie, mckee77, ballet, borkowski, bruggen}. These studies typically find log($N$(\nvp)/$N$(\ovip)) between $-1.1$ and $-1.4$, and log($N$(\civp)/$N$(\ovip)) between $-0.4$ and $-1$ \citep{indebetouw}. Further, these models predict \ovi column densities per velocity interval between 7 and 14~$\times10^{12}$~cm$^{-2}$~km$^{-1}$~s, in agreement with the $C_f$ corrected column densities observed in \autoref{fig:proffits}. While a different physical situation, conductive interfaces also occur in a supernovae driven blastwave and produce similar results \citep{weaver, slavin93, shelton98}. 

\citet{borkowski} model a time-dependent plane-parallel conductive interface between a hot flow and cooler gas. This model computes the column densities of high-ionization lines like \siivp, \civp, \nvp, and \ovi as a function of time, magnetic field orientation, and the physical conditions of the hot wind. In these models the \ovi column density starts with a negligible column density and reaches a saturation level after $\sim$10$^5$~years, which is similar to the observed 0.2~Myr dynamical time of the outflow. Using Equation 4 from \citet{borkowski}, parameters of the hot wind from \citet{chevalier}, and assuming that the magnetic field is parallel to the conduction front; we maximize the total hydrogen in the conductive interface, with a value of $2 \times 10^{19}$~cm$^{-2}$~km$^{-1}$~s. This implies that our observed \ovi is $1\times10^{-4}$ times the total hydrogen, and the total integrated hydrogen column density is $8 \times 10^{21}$~cm$^{-2}$ in the conductive interface. Again, these estimations are upper limits because the values are maximized, but these values allow us to compare upper limits of the transitional gas to the photoionized gas in \autoref{phases}.

\subsubsection{Cooling Flow}
\label{condense}

When hot gas interacts with cool gas, energy flows from the hot gas to the cool gas. If this interaction drops the hot gas temperature to near $5 \times 10^5$~K, the hot gas radiatively cools and rapidly transitions through temperatures corresponding to \ovip.

Recent cooling flow models successfully relate the observed column densities to the line-widths of \ovi and \nv \citep{heckman02, rongmon}. These models predict that the total \ovi column density increases with the line width, approaching values near 10$^{15}$~cm$^{-2}$ for the broadest lines. Using the observed \ovi line width of 650~\kms and the column density of 10$^{15.35}$~cm$^{-2}$, the cooling flow model of \citet{rongmon} predicts the column densities of the other transitions. \autoref{tab:cooling} indicates that the log($N$(\nvp)/$N$(\ovip)) and log($N$(\civp)/$N$(\ovip)) ratios are near $-1.4$ and $-0.7$, respectively. Summing up the total oxygen predicted, we find an \ovi ionization fraction of 0.8\%, a total oxygen column density of $3\times10^{17}$~cm$^{-2}$, and a total hydrogen column density of $4\times10^{20}$~cm$^{-2}$, using the metallicity of the photoionized gas.

\begin{table}
\caption{Estimated column densities ($N$; in units of cm$^{-2}$) of five different ions from the \citet{rongmon} cooling flow model. These values are estimated using an \ovi line width of 650~\kmsp, an \ovi column density of 10$^{15.35}$~cm$^{-2}$, and a log(T) = 5.5 \citep{rongmon}. This model implies that \ovi is 0.8\% of the total oxygen.}
\begin{tabular}{ccccc}
\hline
\siiv & \civ & \nv & \ion{O}{vii} & \ion{O}{viii} \\

[cm$^{-2}$]&[cm$^{-2}$]&[cm$^{-2}$]&[cm$^{-2}$] & [cm$^{-2}$] \\
 \hline
 $7 \times 10^{13}$ & $5 \times 10^{14}$ & $1 \times 10^{14}$ & $7 \times 10^{16}$ & $2 \times 10^{17}$ \\
\end{tabular}
\label{tab:cooling}
\end{table}

\subsection{Which phase dominates the outflow?}
\label{phases}
Now that we have explored the creation of the \ovi gas, we ask: how does the total transitional phase compare to the photoionized phase?  We make this comparison first to the total column densities of the two phases (\autoref{coldencomp}) and then to the mass outflow rates (\autoref{mout}). 

\subsubsection{Comparing the column densities of the two phases}
\label{coldencomp}
The \oi and \ovi column densities are ideal to compare because they arise from the same element, and the only difference between the two transitions is their ionization states. \oi is an unambiguous tracer of neutral gas because it has the same ionization potential as \ion{H}{i}, and the \oi ionization fraction is locked to \ion{H}{i} through charge exchange. The measured \oi column density is 10$^{15.62}$~cm$^{-2}$, nearly a factor of 2 larger than the measured \ovi column density. However, at blue velocities the \ovi column density likely saturates, and the measured \ovi column density could be consistent with the \oi column density.  

The \oi transition only probes neutral gas, which the {\small CLOUDY} model suggests is only 3\% of the total outflow. The wavelength coverage of the \megasaura\ data enables us to calculate the total silicon without assuming a model because it contains transitions from each photoionized silicon phase (\si 1845\AA, \siii 1527\AA, \siiii 1206\AA, \siiv 1402\AA; see \autoref{tab:lines}). Adding up the observed column densities from these four transitions we find a lower limit of the total silicon column density of 10$^{15.33}$~cm$^{-2}$, nearly equal to the total \ovi column density. The silicon abundance in \autoref{cloudy} provides a total photoionized column density of $10^{20.5}$~cm$^{-2}$, as compared to the 10$^{21.9}$ and 10$^{20.6}$~cm$^{-2}$ found for the conductive interface and cooling flow models. Summing the observed silicon transitions implies that the total transitional gas has a column density roughly equal to, and at most 20 times larger than, the column density of the photoionized gas. 
 
However, \siiii is the dominant ion in the photoionized medium and it is heavily saturated. The {\small CLOUDY} model predicts a total hydrogen column density of 10$^{20.75\pm0.04}$~cm$^{-2}$. This suggests that the photoionized column density is 2 and 0.1 times the size of the cooling flow and conductive interface models, respectively. 

Dust along the line-of-sight extincts the observed stellar continuum. Assuming the dust-to-gas ratio scales with metallicity allows the hydrogen column density to be estimated from the measured extinction \citep{claus2002, heckman2011}. Using the relation from \citet{calzetti}, the total hydrogen column density is estimated from the UV extinction as
\begin{equation}
N_\mathrm{H} = \frac{2.4 \times 10^{21} E_\mathrm{gas}(B-V)}{Z} 
\end{equation}
where $E_\mathrm{gas}(B-V)$ is the gas phase extinction. We convert the measured stellar continuum extinction (0.15~mags) to $E_\mathrm{gas}(B-V)$ by dividing by 0.44 \citep{calzetti}. Using the measured extinction and the derived outflow metallicity, we estimate a total hydrogen column density for the photoionized phase of 10$^{20.64 \pm 0.03}$~cm$^{-2}$, where the uncertainty does not account for the extinction law. The inferred column density from the UV extinction is in agreement, within 2$\sigma$, with the total hydrogen column density from the {\small CLOUDY} models. This inferred photoionized column density is twice as large as the cooling flow column density and one-tenth the conductive interface column density. The exact comparison depends on how the \ovi is created and the saturation of \ovip. We qualitatively conclude that the photoionized column density is similar to the cooling flow column densities, and smaller than the conductive interface models. The transitional phase gas is a significant portion of the total outflow, and not including the transitional phase would underestimate the total amount of outflowing gas. 

\subsubsection{Comparing the mass outflow rates of the two phases}
\label{mout}

While deriving column densities does not require assumptions about the outflow geometry, the more fundamental quantity driving galaxy evolution is the mass outflow rate (\moutp). Here, we calculate \mout for both the transitional  ($\dot{M}_\mathrm{tr}$) and photoionized (\mop) phases to compare how the phases contribute to the total \moutp. 

\begin{figure}
\includegraphics[width = 0.5\textwidth]{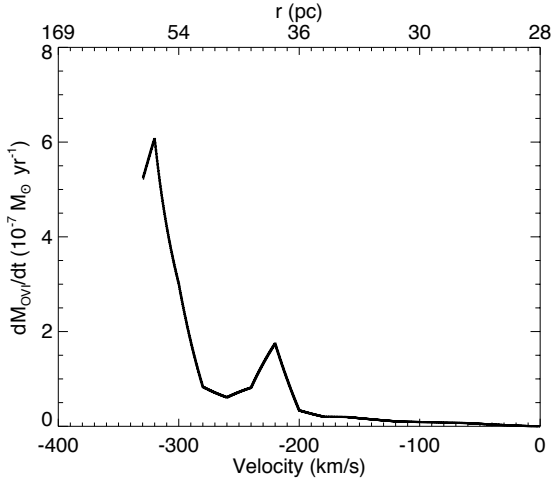}
\caption{The velocity-resolved \ovi mass outflow rate (dM$_\mathrm{OVI}$/dt) in units of 10$^{-7}$~M$_\odot$~yr$^{-1}$. This is calculated assuming that the \ovi and \siiv are co-moving and co-spatial. The \ovi mass outflow rate rises from $0.2 \pm 0.1\times 10^{-7}$~M$_\odot$~yr$^{-1}$ at -180~\kms to $6.1\pm3.0\times10^{-7}$~M$_\odot$~yr$^{-1}$ at -320~\kmsp, increasing by a factor of 30 in 140~\kmsp. At these velocities \ovi is likely saturated and the \ovi mass outflow rate should be considered a lower limit. These are the same velocities that the \alii and \siiv optical depth profiles decline (\autoref{fig:op} and \autoref{fig:proffits}). The upper x-axis gives the radial distance of the outflow from the ionizing source, using the initial radius and the velocity law (\autoref{eq:beta}). }
\label{fig:ovimout}
\end{figure}

We calculate $\dot{M}_\mathrm{tr}$ using the observed \ovi column density and the photoionized line profile fits. This fitting assumes that the \ovi is co-moving with the photoionized phase, as suggested by their line profiles (see the discussion in \autoref{lineprof}). The \ovi mass outflow rate (\mos) is calculated as 
\begin{equation}
\dot{M}_\mathrm{OVI}(\mathrm{r}) = \Omega C_f(r) v(r) m_\textrm{OVI} N_\mathrm{OVI}(r) r
\label{eq:o6moutr}
\end{equation}
Where the radial covering fraction (\autoref{eq:cf}), velocity law (\autoref{eq:beta}), and radius (\autoref{eq:tau0}) are calculated in \autoref{lineprof} (\autoref{tab:lineprof}). Putting these relations into \autoref{eq:o6moutr} gives \mos\ as
\begin{equation}
\dot{M}_\mathrm{OVI}(\mathrm{w}) = \Omega C_f(R_\mathrm{i}) v_\infty m_\textrm{OVI} R_\mathrm{i} N_\mathrm{OVI}(w) \frac{w}{(1-w^{1/\beta})^{1+\gamma}}
\label{eq:o6moutw}
\end{equation}
Where $\mathrm{m}_\mathrm{OVI}$ is 16 times the proton mass (the mass of oxygen atoms), $w$ is the velocity normalized by v$_\infty$ (532~\kms for \ovip), and $\Omega$ is the solid angle covered by the outflow, which we assume is 4$\upi$. We use the $C_f$ corrected \ovi column density distribution ($N_\mathrm{OVI}(w)$) from \autoref{fig:proffits}. Since the \ovi ionization fractions are highly uncertain (see \autoref{o6}), \autoref{fig:ovimout} shows \mos\ (i.e. only the mass outflow rate of the \ovi gas). The \mos\ curve begins at a very low level and steadily increases as the velocity increases. Most of this increase happens between $-180$ and $-320$~\kms where \mos\ increases by a factor of 30. These are the velocities where the \ovi profile is likely saturated (see \autoref{fig:op}), and \mos\ should be considered a lower-limit at these velocities.

\begin{figure}
\includegraphics[width = 0.5\textwidth]{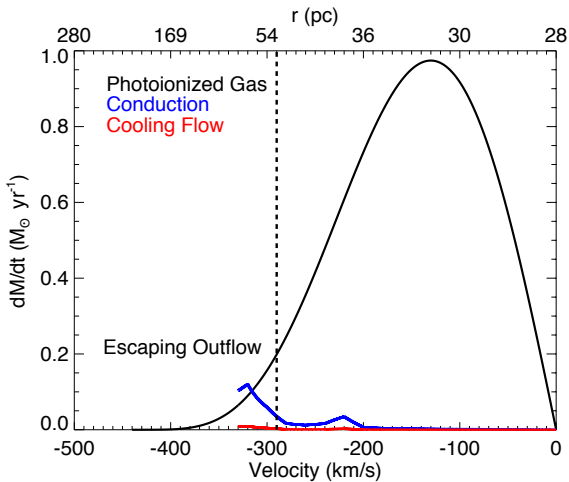}
\caption{The model of the velocity-resolved photoionized mass outflow rate (dM/dt; black line), the mass outflow rate from a conductive interface (blue line), and a cooling flow (red line). All of these estimates are for the total hydrogen mass outflow rate in their respective phases. Both the blue and red lines should be treated as upper limits due to uncertain ionization corrections. The photoionized mass outflow rate peaks at $0.97\pm0.39$~\sfrp. The maximum mass outflow rate of the conductive interface model is 12\% the peak of the photoionized gas. At $-290$~\kmsp, a dashed line represents the upper-limit of the escape velocity for J1226+2152. Any outflow blueward of this velocity is traveling fast enough to escape the gravitational potential of the galaxy. The photoionized outflow rapidly decreases at velocities above the escape velocity, while the transitional mass outflow rate dramatically increases and actually exceeds the photoionized outflow. The upper x-axis gives the radial distance of the outflow from the ionizing source, using the initial radius and the velocity law (\autoref{eq:beta}).}
\label{fig:moutcomp}
\end{figure}

Using the ionization corrections found in \autoref{o6}, we derive upper-limits of the total $\dot{M}_\mathrm{tr}$. $\dot{M}_\mathrm{tr}$ peaks at $-320$~\kms with values of 0.12~\sfr and 0.009~\sfr for the conductive interface and cooling flow models, respectively (\autoref{fig:moutcomp}). These estimates are for the total hydrogen mass outflow rate within the conductive interfaces or cooling flows, but do not include the mass in a hotter phase. It is important to stress that $\dot{M}_\mathrm{tr}$ likely continues rising at the bluest velocities, but the covering fraction corrected \ovi column density is not defined past $-340$~\kms because the \siiv profile is no longer detected (see \autoref{fig:proffits}).

We follow \citet{chisholm16b} and calculate \mop\ as
\begin{equation}
\begin{aligned}
    \dot{M}_\mathrm{ph}(r) &= \Omega C_{f}({r}) {v}(r) \rho(r) r^2 \\
    \dot{M}_\mathrm{ph}(w)&= \Omega C_f({R}_\mathrm{i}) {v}_\infty \mu \mathrm{m}_\mathrm{p}  {n}_\mathrm{H,0} {R}_\mathrm{i}^2 \frac{w}{(1-w^{1/\beta})^{2+\gamma+\alpha}}
    \label{eq:mout}
\end{aligned}
\end{equation}
Now we use 1.4 times the mass of the proton (the effective mass per nucleon; $\mu\mathrm{m}_\mathrm{p}$) because we are calculating the total outflowing mass rather than just the \ovi mass (we also use this value when calculating $\dot{M}_\mathrm{tr}$ for the conductive interface and cooling flow models above). The parameters are taken from the photoionization modeling and the \siiv profile fitting, where $v_\infty$ is $-440$~\kms for the \siiv line. $n_\mathrm{H,0}$ in \autoref{eq:mout} is the total hydrogen density from the photoionization modeling in \autoref{cloudy}. Importantly, \autoref{eq:mout} has the exact same $C_f(r)$, $v(r)$, and $R_i$ as \autoref{eq:o6moutr}. The difference between the two relations is the radial density distribution ($n(r)$ versus $N_\mathrm{OVI}$) and the maximum velocity. 

The maximum \mop\ peaks at $1.0\pm0.4$~\sfr and decreases at higher and lower velocities, where the \mop\ errors are calculated using a Monte Carlo analysis of the input parameters from \autoref{eq:mout}. The conductive interface \mout peaks at -320~\kmsp, with an upper limit of 0.12~M$_\odot$~yr$^{-1}$ (blue line in \autoref{fig:moutcomp}), only 12\% of the peak of \mop. The cooling flow model peaks at the same velocity with an upper limit of 0.009~\sfr (red line in \autoref{fig:moutcomp}). The photoionized outflow is the chief contributor to \mout at the observed velocities. 

$\dot{M}_\mathrm{tr}$ rapidly rises at velocities blueward of $-200$~\kmsp, reminiscent of the velocity that splits the \ovi profile into two regimes. The  $\dot{M}_\mathrm{tr}$ rises precisely at the same velocity as \mop\ declines. In fact, at $-320$~\kms the \mout of the conductive interface model exceeds \mop, indicating that $\dot{M}_\mathrm{tr}$ may dominate at bluer velocities. The continued acceleration of the transitional gas, saturation of the \ovip, and the presence of a hotter undetected wind may further increase $\dot{M}_\mathrm{tr}$ and provide the additional 0.9~\sfr required to satisfy mass conservation. These requirements may indicate that the undetected hot wind component dominates the mass outflow rate at higher velocities. 

The high velocity gas is the most important for galaxy evolution because this gas escapes the gravitational potential of the galaxy. \citet{wuyts} place an upper-limit on the stellar mass of J1226+2152 at log(\mstarp)$~<~9.5$, which leads to an upper limit on the circular velocity of $<95$~\kms \citep[using the Tully-Fischer relation from][]{reyes}. \citet{heckman2000} approximate the escape velocity as three times the circular velocity, which means that an upper-limit for the escape velocity is $<-290$~\kmsp. At these velocities \mop\ rapidly declines while $\dot{M}_\mathrm{tr}$ increases; outflows above the escape velocity may entirely consist of hot and transitional gas. This likely means that studies focusing on cool photoionized gas to calculate the amount of mass escaping galactic potentials under-predict the total mass of high-velocity outflows. An analysis of the hotter phases are required to accurately determine this quantity.

\subsection{A physical model explaining these observations}
\label{model}
Here we synthesize the results of the past sections into a physical model that could explain the observations. First, we summarize the observations that our model tries to explain: 
\begin{enumerate}
    \item The \ovi line profile has two regimes, where it matches weak lines at redder velocities and strong lines at bluer velocities (see \autoref{fig:lineprof} and \autoref{fig:op}). Changes in covering fraction drive changes in the line profiles of strong lines, while changes in optical depth drive changes in the line profiles of weak lines. The division line for the \ovi profile happens approximately at $-200$~\kmsp. 
    \item The covering fraction corrected \siiv column density drops sharply at bluer velocities (\autoref{fig:proffits}).
    \item The covering fraction corrected \ovi column density is flat at low velocities and increases at high velocities (\autoref{fig:proffits}). The \ovi profile transitions from being optical depth dominated at low velocities to covering fraction dominated at high velocities. The \ovi column density starts increasing at $-200$~\kmsp. 
    \item \mop\ increases at low velocity and decreases at higher velocities as the density and covering fraction decline. As \mop\ decreases, \mos\ increases by a factor of 30 (\autoref{fig:ovimout}) and actually exceeds \mop\ at the highest observed velocities (\autoref{fig:moutcomp}).
    \item The \ovi and photoionized gas are not created by the same mechanism. The {\small CLOUDY} models do not reproduce the observed \ovi column densities (\autoref{tab:lines}), but reasonably reproduce the photoionized gas. The \nv non-detection indicates that the \ovi is likely produced through conductive evaporation of the photoionized gas or through a cooling flow of a hotter outflow (\autoref{o6}).
\end{enumerate}

\begin{figure}
\includegraphics[width = 0.5\textwidth]{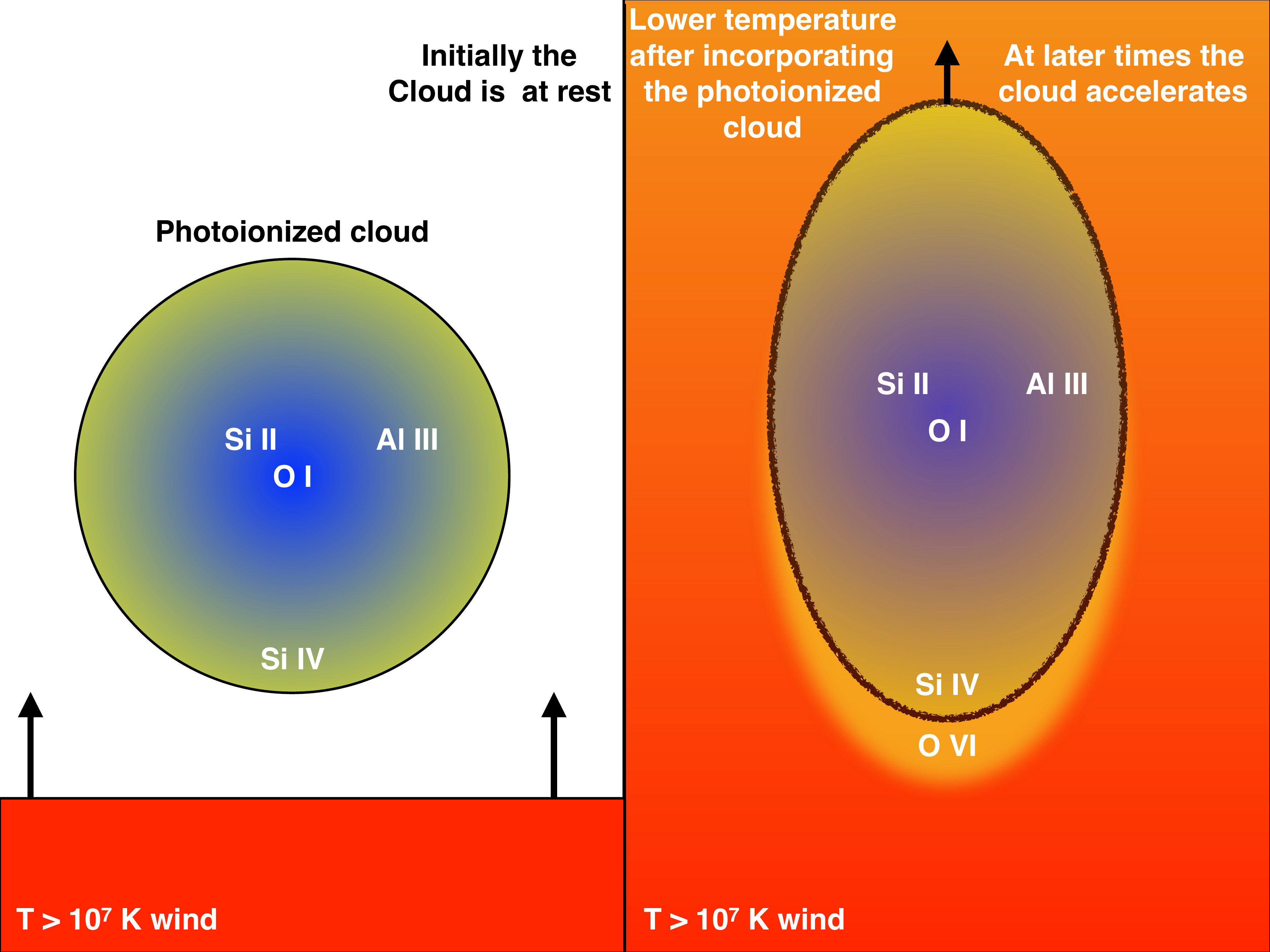}
\caption{Cartoon of the physical picture that produces the \ovi absorption. {\it Left Panel:} Initial conditions of a photoionized gas cloud at rest. A nearby starburst (not pictured) created a $T>10^7$~K wind that propagates towards the photoionized cloud. Hypothetical locations of various observed ions are labeled. {\it Right Panel:} A time after the hot wind encountered the photoionized cloud. Ram pressure accelerates and elongates the cloud, while conduction between the hot outflow and the photoionized cloud evaporates the cloud to produce \ovi at the interface. Energy from the hot outflow further heats and accelerates the \ovip, which reduces the down-stream temperature of the hot wind. By incorporating the photoionized cloud, the down-stream outflow has an increased mass, but a lower temperature, velocity, and metallicity.}
\label{fig:ovischem}
\end{figure}

Here we envision a simple model where a hot (>10$^{7}$~K) wind is incident on an ensemble of photoionized clouds (see a zoom-in of a single cloud in \autoref{fig:ovischem}). The hot outflow accelerates the photoionized clouds from zero velocity to high velocities by ram pressure, as described by the velocity law in \autoref{proffits}. At the interface of each cloud, heat transfers from the hot outflow to the photoionized clouds through conduction \citep{weaver}, heating the outer layer of photoionized gas to high temperatures. The mass transfer reduces the density of the photoionized gas, while increasing the column density of the \ovi and hot wind (lower right panel of \autoref{proffits}). Geometric dilution, elongation from the ram pressure shock, and adiabatic expansion reduce the size and covering fraction of the photoionized clouds \citep{klein, martin09, scannapieco15, chisholm16b, bruggen}. After about 0.2~Myr (the dynamical time of the photoionized clouds) conduction has evaporated most of the photoionized gas and incorporated it into the hot wind \citep{bruggen}. The process increases the mass of the hot wind, while decreasing its average metallicity, temperature, and velocity \citep{thompson16}. The transfer of mass from the photoionized phase to the hot wind is often referred to as the mass-loading of the hot wind \citep{maclow, suchkov, heckman2000, strickland2000, cooper09, strickland09}.

This model explains the observations in a number of ways. First, the \ovi is co-spatial with the photoionized gas because the \ovi only temporarily arises in the outer layer of the photoionized clouds (see \autoref{fig:ovischem}). Second, to satisfy mass conservation, the photoionized mass must directly transfer to the hot wind. Therefore, as the photoionized density and \mop\ decreases, $\dot{M}_\mathrm{tr}$ and $\dot{M}_\mathrm{hot}$ must increase. This indicates that $\dot{M}_\mathrm{hot}$ should be substaintial at blue velocities.

Mass-loading of a hot wind with photoionized gas naturally explains the observed two-regime \ovi line profile from \autoref{lineprof}. Before the hot wind interacts with the photoionized cloud, the cloud is at rest and the photoionized density is large (left panel of \autoref{fig:ovischem}). Since photoionization produces negligible \ovip, the \ovi is optically thin at redder velocities. However, once the hot wind interacts with the photoionized cloud, conduction heats the cloud to temperatures briefly corresponding to \ovi (right panel of \autoref{fig:ovischem}). As the interaction continues, conduction forges more \ovi by heating more photoionized gas. Consequently, while the hot wind accelerates the cloud, the weak low-ionization lines become optically thin as the \ovi becomes optically thick. This is precisely what we observe in the lower right panel of \autoref{fig:proffits}. Once \ovi saturates, the  $C_f$ of the clouds determines the shape of the profile. This is why the \ovi profile closely follows the shape of other saturated lines (\siii and \ciip). Feeding photoionized gas into a hot wind creates the two-regime \ovi line profile because at low velocities conduction has not acted long enough to produce appreciable \ovip, while at high velocities conduction has produced enough \ovi to saturate the transition.

A prediction of this model is that as the outflow becomes more ionized (and consequently less dense), the velocity must increase to satisfy the continuity equation. The increased velocity causes the mass outflow rate of increasing ionization states to peak at bluer velocities (from photoionized gas to transitional gas in \autoref{fig:moutcomp}). \citet{chisholmmat} find that cold molecular outflows have lower outflow velocities than neutral outflows, and this model predicts that, if J1226+2152 has a molecular outflow, the molecular mass outflow rate should peak at velocities redward of $-130$~\kmsp.

\subsubsection{Does mass-loading of a hot wind impact the CGM?}
\label{cgm}
The small physical distance and the rapid destruction of photoionized clouds means that the observed photoionized clouds do not survive to travel into the circum-galactic medium (CGM). However, cool gas is observed in the CGM around galaxies \citep{tumlinson, werk,bordoloi14, borthakur15, prochaska17}. Here, we explore reconciling this issue by reforming the warm CGM through a radiative shock of the hot, mass-loaded wind \citep{wang95, wangb, thompson16}, and the implications that the proposed model of mass-loading has on the formation of the CGM.

At $1000-3000$~\kms \citep{chevalier, bustard}, the $T>10^7$~K hot wind contains most of the outflowing energy. The hot wind travels out of the star-forming region where ambient gas is incorporated into the hot wind. After it is mass-loaded, the hot phase travels outward with a reduced velocity because the acceleration and heating of the ambient gas consumes energy from the hot wind. Gravity, inter-galactic medium pressure, radiative cooling, and adiabatic expansion further decreases the velocity and temperature of the hot outflow until it reaches the peak of the cooling curve, which causes the hot outflow to radiatively shock and reform cool photoionized gas \citep{wang95, wangb, thompson16}. 

Changing the initial composition, density, and temperature of the hot wind changes where the hot wind radiatively shocks. \citet{thompson16} predict that the cooling radius of the hot wind scales with the mass outflow rate and metallicity of the hot wind as $\dot{\mathrm{M}}_\textrm{hot}^{-2.92}$ and Z$_\mathrm{hot}^{-0.79}$, respectively. If we assume that the hot wind initially has the mass outflow rate of the transitional gas in \autoref{mout} (1/8th the photoionized outflow rate) with the metallicity of pure supernovae ejecta \citep[$\sim5$~Z$_\odot$;][]{woosley}, the addition of photoionized mass and metals will increase $\dot{\mathrm{M}}_\textrm{hot}$ but decrease Z$_\mathrm{hot}$, decreasing the cooling radius by a factor of 200 from 10~Mpc scales to 100~kpc scales. Without the mass-loading of cool clouds into the hot wind, the hot wind would not be massive enough to produce the observed metal-enriched CGM within 100~kpc of galaxies \citep{werk, prochaska17}. 

Further, whether the hot outflow shocks at all depends on the mass-loading. \citet{thompson16} show that there is a minimum ratio between \mout and SFR--the so called mass-loading factor--above which the hot wind will radiatively shock. If the mass-loading factor is lower than this minimum mass-loading factor, the hot wind does not lose enough thermal energy to reach the peak of the cooling curve, and never recreates the low-ionization gas observed in the CGM. \citet{thompson16} find that the minimum mass-loading factor ranges between \moutp/SFR$=$0.3$-$0.6. 

Since the mass-loading factor of galactic outflows decreases with increasing stellar mass \citep{rupke2005b, heckman2015, chisholm17}, galaxies greater than $2-8\times10^{10}$~M$_\odot$ generate outflows with mass-loading factors below the minimum mass-loading factor \citep{chisholm17}. Hot winds in high-mass galaxies never radiatively shock to produce \ovi and low-ionization gas in the CGM. Instead the temperature of the hot wind remains above the peak of the cooling curve. \citet{tumlinson} detect \ovi absorption in 100\% of the 19 galaxies with stellar mass less $4\times10^{10}$~M$_\odot$, but this detection rate drops to 52\% for the 23 galaxies with stellar mass greater than $5\times10^{10}$~M$_\odot$ (note that these high-mass galaxies contain both star-forming and quiescent galaxies). Similarly, other studies find that the majority of \ovi absorption in the CGM arises from galaxies that are fainter than L$^\ast$ \citep{tumlinson2005, stocke, Prochaska2011b, pratt}. Since \ovi has a cooling time of approximately 1~Myr \citep{gnat}, it acts as a clock for the formation of the CGM. This indicates that the low-ionization gas in the CGM of high-mass galaxies is a relic of an earlier, more mass-loaded, galactic outflow. The mass-loading of a 10$^7$~K wind with cooler photoionized gas may enable the formation of the observed CGM.

\section{SUMMARY}
We searched for \ovi in the six galaxies with far-ultraviolet coverage from \megasaura.  We found \ovi in 3 of the 4 spectra with signal-to-noise ratios greater than 3 (\autoref{fig:megasaura}). We then analyzed the interstellar absorption features in the spectrum of SGAS~J122651.3+215220 (J1226+2152), a galaxy at a redshift of $2.926\pm0.0002$ that has the highest signal-to-noise ratio near the \ovi~1032\AA\ interstellar absorption line. We used the \ovi line, which probes gas transitioning between warm (10$^4$~K) and hot (>10$^7$~K) temperatures, to study a seldom-probed phase of galactic outflows. The \ovi has an observed column density of 10$^{15.35\pm0.05}$~cm$^{-2}$ (although it is likely saturated at the bluest velocities) and a maximum outflow velocity near -530~\kmsp. The maximum velocity is consistent with saturated low-ionization tracers like \ciip~1335\AA\ and \siiip~1260\AA\ (\autoref{tab:lines}). We found that the \ovi line profile has two regimes: a low-velocity portion that closely resembles weaker low-ionization lines, and a high-velocity portion that resembles the strong saturated lines. The \ovi line profile transitions between these two regimes at a velocity of $-200$~\kms (\autoref{fig:lineprof} and \autoref{fig:op}). The \ovi and photoionized gas have similar velocity profiles and covering fractions, indicating that the two phases are co-spatial. 

Using the \siiv doublet, we studied how the optical depth and covering fraction of low-ionization gas varies with velocity. We then corrected the observed \siiv column density for the covering fraction and found that the \siiv column density drops at velocities blueward of -200~\kmsp. After correcting the \ovi column density for the \siiv covering fraction, the \ovi column density rises at velocities blueward of -200~\kmsp. \ovi is produced as the low-ionization gas is destroyed (\autoref{fig:proffits}). 

We used five weak absorption lines to model the ionization structure of the outflow with {\small CLOUDY} photoionization models. We found that many of the low-ionization lines are adequately predicted by the photoionization models, but the models produce negligible \ovi (see columns 6 and 7 in \autoref{tab:lines}). Combined with the non-detection of the \nv line (\autoref{fig:nvlineprof}), we postulated that the \ovi is created either through conductive evaporation of the photoionized gas by a hot outflow, or through a cooling flow between a hot outflow and the cooler photoionized gas. We used these two models to derive upper limits of the \ovi ionization corrections (\autoref{o6}).

We calculated upper limits of the \ovi mass outflow rate and found that it steadily rises at blue velocities (see \autoref{fig:ovimout}).  The maximum transitional mass outflow rate is at least 10-100 times lower than the maximum of the photoionized mass outflow rate, depending on the \ovi ionization mechanism and saturation of \ovip. This suggests that the photoionized gas dominates the mass outflow rate at the observed velocities, although the mass outflow rate of the transitional phase rises, and overtakes, the photoionized mass outflow rate at the bluest velocities (\autoref{fig:moutcomp}). 

The \ovi phase likely dominates the mass outflow rate at velocities above the escape velocity (\autoref{fig:moutcomp}). This high-velocity gas completely removes mass from the galaxy, however it is not typically probed by rest-frame ultraviolet and optical observations. If studies want to measure the amount of gas removed from galaxies, the hotter outflow phases must be considered.  

Finally, we put forth a physical picture where the observed \ovi traces the destruction of the photoionized outflow by conduction and the incorporation of the photoionized gas into an unobserved $>10^7$~K wind (\autoref{fig:ovischem}, \autoref{model}). To satisfy conservation of mass, the sum of the mass outflow rate for all of the different phases must remain constant, implying that the mass outflow rate of the transitional and hot wind increases as the photoionized mass outflow rate decreases. This is exactly what we observed in \autoref{fig:moutcomp}. This also implies that the mass outflow rate of the unobserved hot wind is massive and increasing at the largest velocities. We speculated on the effect of incorporating low-ionization gas into a hot wind and the formation of the circum-galactic medium (\autoref{cgm}). 

These observations provide new evidence for which gas escapes the gravitational potentials of galaxies. The high-velocity outflow drives the depletion of gas within galaxies, and quantifying it is important to understand what drives the star formation histories of galaxies. Larger samples of rest-frame ultraviolet spectra of galaxies with $z > 2.5$ will constrain the amount of gas removed from galaxies by hot outflows. J1226+2152 is one of the brightest gravitationally lensed galaxies at these redshifts, and it required 12.4~hr of integration time on the 6.5~m Magellan telescope. Larger surveys must wait for $20-30$~m class telescopes. Furthermore, actually probing the elusive 10$^7$~K wind phase is the largest, and possibly most important, missing component of galactic outflows. Future X-ray observatories, with higher sensitivity and spectral resolution, will measure the hot wind in galaxies outside of the local universe, determining whether hot winds contain the majority of the outflowing mass that escapes galaxies. 

\section*{Acknowledgments}
We thank the referee for a thoughtful read of the paper, and comments that strengthened the work. 

Support for this work was provided by NASA through Hubble Fellowship grant \#51354 awarded by the Space Telescope Science Institute, which is operated by the Association of Universities for Research in Astronomy, Inc., for NASA, under contract NAS 5-26555.

\bsp	
\label{lastpage}

\bibliographystyle{mnras}
\bibliography{o6}

\end{document}